\documentclass[12pt]{article}
\usepackage{amsmath,amsfonts,amssymb,amscd,xcolor,ifpdf}

\usepackage{epsfig}

\newcommand{\lj}{\mbox{$[\kern-0.1478125em[$}}
\newcommand{\rj}{\mbox{$]\kern-0.1478125em]$}}

\newcommand{\BFX}{{\bf x}}
\newcommand{\BFB}{{\bf B}}
\newcommand{\BFJ}{{\bf J}}

\newcommand{\BFT}{{\bf t}}

\newcommand{\BFA}{{\bf A}}

\newcommand{\be}{\begin{equation}}
\newcommand{\ee}{\end{equation}}

\newcommand{\BFx}{\widehat{{\bf x}}}

\newcommand{\BFy}{\widehat{{\bf y}}}
\newcommand{\BFz}{\widehat{{\bf z}}}

\newcommand{\Fig}[1]{Fig.~\ref{#1}}
\newcommand{\Eqn}[1]{Eq.~(\ref{eq#1})}

\title{Elimination of MHD current sheets by modifications to the plasma wall in a fixed boundary model}
\author{E. Kim,$^1$ G.B. McFadden,$^2$ and A.J. Cerfon$^1$}
\date{}
\begin{document}
\maketitle

\begin{center}

$~^1$Courant Institute of Mathematical Sciences, NYU, New York, NY 10012 \\
$~^2$Applied and Computational Mathematics Division, NIST, Gaithersburg, MD 20879 \\

\date{\today}

\end{center}

\begin{abstract}
Models of magnetohydrodynamic (MHD) equilibia that for computational convenience assume the existence of
a system of nested magnetic flux surfaces tend to exhibit singular current sheets. These sheets are located on
resonant flux surfaces that are associated with rational values of the rotational transform. We study the
possibility of eliminating these singularities by suitable modifications of the plasma boundary, which 
we prescribe in a fixed boundary setting. We find that relatively straightforward iterative procedures can be
used to eliminate weak current sheets that are generated at resonant flux surfaces by the nonlinear interactions
of resonating wall harmonics. These types of procedures may prove useful in the design of fusion devices with
configurations that enjoy improved stability and transport properties.
\end{abstract}


\section{Introduction}

We describe a modification of the magnetohydrodynamic (MHD) equilibrium and
stability code NSTAB \cite{Tayl94} in order to study the effect of wall perturbations
on resonant flux surfaces where singular current sheets are often observed.
NSTAB (and our modified version of NSTAB) 
solves the governing equations
\be \label{eq1}
     \BFJ \times \BFB = \nabla p, \quad \BFJ = \nabla \times \BFB, \quad \nabla \cdot \BFB = 0,
\ee
where $\BFB$ is the magnetic
field, $\BFJ$ is the current density, $p$ is the plasma pressure, and to simplify the notation we normalized the magnetic permeability to unity. The first equation in (\ref{eq1}) is the basic
force balance in the plasma, the second defines the current density, and the
third expresses the solenoidal nature of the magnetic field.

NSTAB is a fixed boundary code, meaning that the plasma boundary $\partial V$ of the toroid plasma volume $V$ is considered to be given with a prescribed shape. NSTAB equilibria are stationary points of
the energy functional \cite{Grad66,KrKu58,Tayl94}
\be \label{eq2}
     E =  \int_V \left\{ \frac{|\BFB|^2}{2} - p \right\} \, dV,
\ee
which is extremized over solenoidal fields $\BFB$ that have
vanishing normal flux at the boundary of $V$. Following \cite{BaBG78,BaBG84,BBGW87,HiWh83,Tayl94}, 
an assumption of {\it nested flux surfaces} is used in
formulating the model, and the problem is recast as a variational principle over a class of
functions that satisfy this constraint. Although the assumption of nested flux surfaces provides
significant computational advantages, the price to be paid is that this assumption can introduce
singularities in the solution \cite{Gara99,BeMc88,Huds00,Nire19}, as will be discussed in some detail.

To implement the constraint of nested flux surfaces the toroidal flux itself is introduced as an independent variable,
and the variational principle is posed  in a fixed computational domain that is defined in terms of the independent
variables $(s,u,v)$ in a unit cube. Here $s$ is the normalized toroidal flux
with $0 < s < 1$, and $u$ and $v$ are normalized poloidal and toroidal angles. Since $\BFB \cdot \nabla p = 0$,
the pressure is constant on an ergodic flux surface, and the problem formulation is
completed by prescribing the pressure field $p(s)$
and the rotational transform $\iota(s)$ \cite{Bate80} as functions of the flux label $s$.

\begin{figure}[h]
\begin{center}
\resizebox{3.0in}{!}{\includegraphics{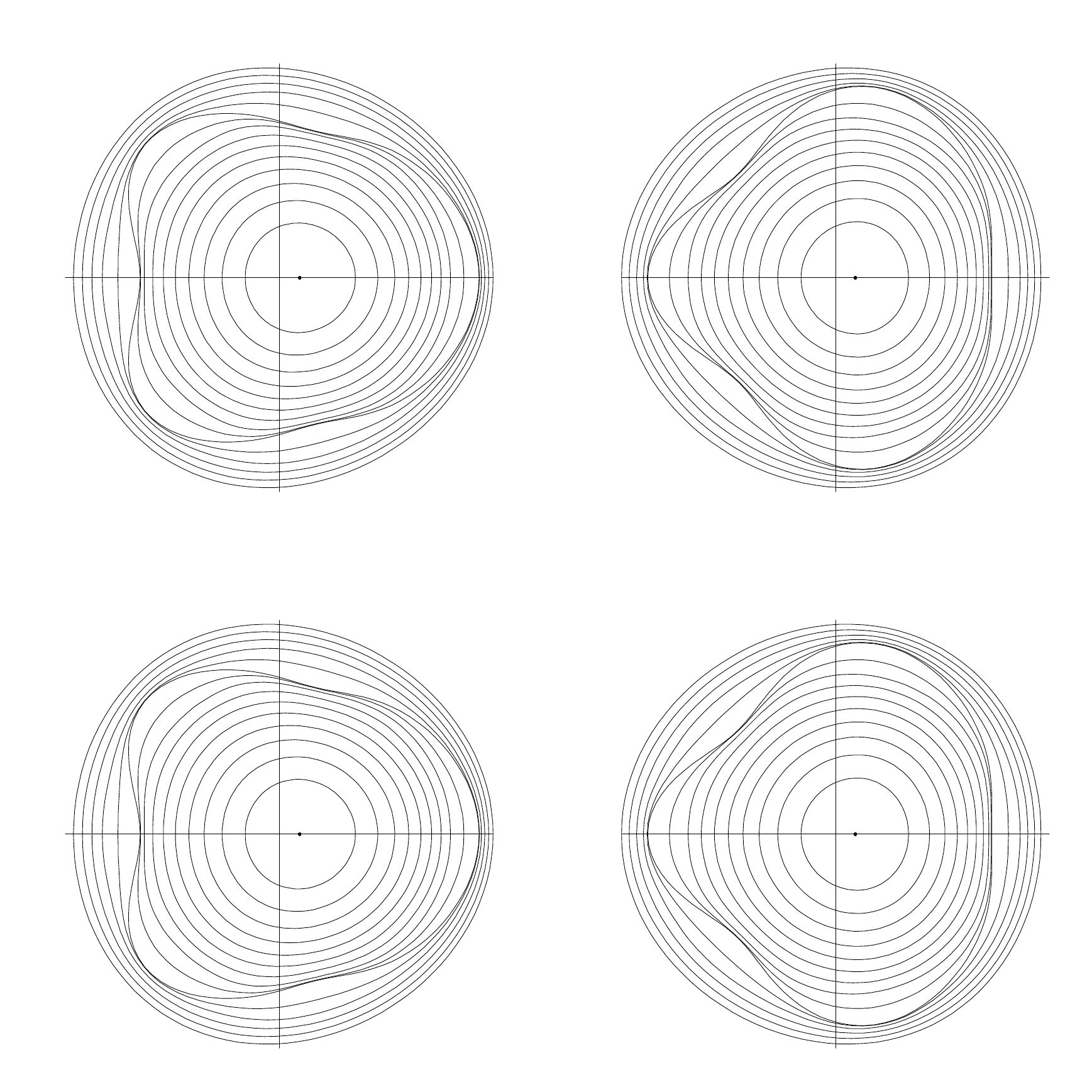}}
\parbox{6.0in}{
\caption{NSTAB computation of four cross-sections of a
torus with aspect ratio $A = 5$. From left to right and top to bottom,
the normalized toroidal angles are $v = 0$,
$v = 1/4$, $v = 1/2$ and $v = 3/4$. The
pressure is given by $p(s) = 0.01 (1 - s^2)^2$, and the rotational transform
is $\iota(s) = 0.6 + 0.1 s$. The observed spacing
of the displayed flux surfaces (contours of constant $s$) reflects a
singular current sheet that occurs at a resonant flux surface where $\iota(s) = 2/3$,
and the surfaces display a corresponding symmetry with poloidal and
toroidal mode numbers $(m,n) = (3,2)$. }
\label{fig0} }
\end{center}
\end{figure}

The singularities in this model 
tend to occur on so-called {\it resonant flux surfaces} where the rotational transform assumes rational values, $\iota(s) = n/m$.
These singularities can be interpreted as current sheets located at the resonant flux surfaces which are present to prevent the formation of islands that would otherwise develop in an equilibrium without the constraint of nested flux surfaces \cite{Huds00}.
An example of an NSTAB calculation with current sheets is shown in Fig.~1,
where the flux surfaces alternately bunch up and spread out around a resonant surface where $\iota(s) = 2/3$.
The $(3,2)$ symmetry of the resonance is clear in the cross-sections of the flux surfaces at four stations
around the torus, and the local distortions occur at points where the Jacobian $\partial(x,y,z)/\partial(s,u,v)$
of the mapping from computational to physical coordinates assumes large or small values. These local distortions are representative of the tendency of the equilibrium to allow for magnetic islands, if the constraint of nested flux surfaces were relaxed. The relation
between the singular current sheet, as reflected in the behavior of the mapping Jacobian, and the prescribed wall
geometry is the subject of
this study, which will be conducted numerically in a simplified geometry. A detailed theoretical discussion of the underlying
differential equations requires the machinery of KAM theory \cite{KAM} which hinges upon the occurrence of ``small divisors'' in the
problem. These present significant numerical issues for convergence under mesh refinement; here we will confine our
attention to the numerical treatment of the underlying discretized model in which the problem of small divisors is sidestepped by
restricting the degrees of freedom in the angular coordinates \cite{Gara99}.

Desirable MHD equilibria with good particle confinement typically feature a large fraction of the plasma volume with nested flux surfaces, and few, narrow magnetic islands with correspondingly small regions with magnetic field stochasticity surrounding them. A corresponding NSTAB equilibrium would exhibit only weak current sheets at resonant surfaces. In this paper, we explore the possibility of prescribing the shape of the fixed plasma boundary in such a way
that resonant singularities are suppressed. To do so, we have examined a modified form of NSTAB that is suitable for a slab geometry, in a doubly-periodic domain that is bounded by two given flux surfaces. We chose to consider this simplified geometry, corresponding to a topological torus without the curvature effects of a true torus, in order to avoid complications associated with the magnetic axis \cite{BaBG78,BaBG84,BBGW87,Tayl94,Weitzner14}
that is surrounded by the innermost flux surface in a toroidal geometry, which represents a coordinate singularity
requiring special numerical treatment. Our work is motivated by the recent analysis of Weitzner \cite{Weitzner14}, which suggests that one can tailor the outermost flux surface in order to avoid resonance-induced singularities. The present study shares many similarities with the recent work of Mikhailov, N\"uhrenberg and Zille \cite{Nire19}, with the following notable differences. Mikhailov \textit{et al.} remove singularities in true toroidal stellarator equilibria computed with VMEC \cite{HiWh83}, as opposed to the slab geometry in the present work. On the other hand, we will show that we can adjust the outermost flux surface in order to remove singularities at multiple resonant flux surfaces, whereas Mikhailov \textit{et al.} only focused on the removal of a single singularity. In addition, we demonstrate that the method also applies to equilibria with pressure profiles with a finite pressure gradient throughout the plasma volume, whereas in their work based on VMEC equilibria, Mikhailov \textit{et al.} flattened the pressure profile in the neighborhood of the resonant flux surface before removing the current sheet through an appropriate boundary perturbation.

The structure of this article is as follows. In the next section, we present the formulation for the slab version of NSTAB, which we have called NSLAB. In the following section, we derive a linearized model starting from the NSLAB formulation, in which the singularity at resonant flux surfaces appears explicitly. We then present numerical results for the full set of nonlinear governing equations, and end the article with some conclusions and suggestions for future work.

\section{A slab version of NSTAB: NSLAB}
\label{sec:NSTAB}

\subsection{Governing equations}\label{sec:NSTAB_Equations}
We describe a modification of the stellarator equilibrium code NSTAB \cite{Tayl94}, denoted ``NSLAB,'' 
that solves the governing equations (\ref{eq1})
in a topological torus, or slab geometry, allowing us to avoid dealing with the
magnetic axis that occurs in a toroidal geometry.
The physical domain is assumed to be doubly periodic in $x$ and $y$, which play the role of the ``poloidal'' and
``toroidal'' angles in this simplified geometry. The fields have periods $L_x$ and $L_y$,
with $x = L_x u$ and $y = L_y v$, where $0 \le u \le 1$ and
$0 \le v \le 1$. The mapping to
physical space $\BFX = \BFX(s,u,v)$ 
is given by
\be
     x = L_x u, \quad y = L_y v, \quad z = z_0(u,v) + R(s,u,v) [z_1(u,v) - z_0(u,v)],
\ee
where $z_0(u,v)$ and $z_1(u,v)$ are the coordinates of the lower and upper flux surfaces,
corresponding to $s = 0$ and $s = 1$, respectively, with $R(0,u,v) = 0$ and $R(1,u,v) = 1$, and where we follow the dimensionless treatment of the governing equations used in NSTAB \cite{Tayl94}, wherein the characteristic length scale is given by the minor radius.
The monotonicity of $R(s,u,v)$ as a function of $s$ incorporates the
assumed constraint of nested flux surfaces. In physical space, the flux surfaces are the graphs of the 
function $z(s,u,v)$ as a function of $u$ and $v$ for constant $s$.

The solenoidal magnetic field $\BFB(s,u,v)$ is represented in terms of a Clebsch potential $\psi(s,u,v)$ 
as
\be \label{eq4}
     \BFB = \nabla \psi \times \nabla s =  B^u \BFT_u + B^v \BFT_v,
\ee
where the contravariant basis vectors are
\begin{align}
    \BFT_u = & \, \frac{\partial \BFX}{\partial u} = L_x \BFx + \left[] \frac{\partial z_0}{\partial u} +
          R \left( \frac{\partial z_1}{\partial u} - 
           \frac{\partial z_0}{\partial u} \right) +
    (z_1 - z_0) \, \frac{\partial R}{\partial u} \right] \BFz, \\
     \BFT_v = & \, \frac{\partial \BFX}{\partial v} = L_y \BFy + \left[] \frac{\partial z_0}{\partial v}  +
          R \left( \frac{\partial z_1}{\partial v} - 
           \frac{\partial z_0}{\partial v} \right) +
    (z_1 - z_0) \, \frac{\partial R}{\partial v} \right] \, \BFz,  \\
   \BFT_s = & \, \frac{\partial \BFX}{\partial s} = (z_1 - z_0) \, \frac{\partial R}{\partial s} \, \BFz.
\end{align}
Here $\BFx$, $\BFy$, and $\BFz$ are unit vectors in the $x$, $y$, and $z$ directions.
The Jacobian of the coordinate transformation is given by
\be \label{eq:jac}
   J(s,u,v) = \frac{\partial(x,y,z)}{\partial(s,u,v)} = \BFT_s \cdot \BFT_u \times \BFT_v =
   L_x \, L_y \, [z_1(u,v) - z_0(u,v)] \, \frac{\partial R}{\partial s}(s,u,v),
\ee
and the contravariant components of $\BFB$ are
\be
    B^u(s,u,v) = \frac{1}{J} \frac{\partial \psi}{\partial v}(s,u,v), \quad    
    B^v(s,u,v) = -\frac{1}{J} \frac{\partial \psi}{\partial u}(s,u,v).
\ee
We observe that since \Eqn{4} may also be written
as $\BFB = \nabla \times (\psi \nabla s)$, we may interpret $\psi$ as the $s$-component 
of a covariant vector potential $\BFA = \psi \nabla s$, so that the corresponding 
current density is $\BFJ = \nabla \times [\nabla \times (\psi \nabla s)]$.

With this representation, the dependent variables in NSLAB are $R(s,u,v)$ and $\psi(s,u,v)$, which satisfy partial differential equations that
result from the the first variation of the energy (\ref{eq2}),
\begin{align} 
 0 = \delta E = \iiint \left[ L_1(\psi) \delta \psi + L_2(R) \delta R \right] \; \mathrm{d}s\, \mathrm{d}u\, \mathrm{d}v.
\end{align}
The Euler-Lagrange equations $L_1(\psi) = 0$ and $L_2(R) = 0$ 
can be written in the form \cite{Tayl94}
\begin{align}
 L_1(\psi)  = &  \, \frac{\partial B_u}{\partial v} - \frac{\partial B_v}{\partial u} \label{eq13a} = 0\\
 L_2(R)     = & \,\frac{\partial \psi }{\partial u}\, \left[ \frac{\partial B_s}{\partial v}
        - \frac{\partial B_v}{\partial s} \right] -\frac{\partial\psi}{\partial v} \,\left[\frac{\partial B_s}{\partial u}
 - \frac{\partial B_u}{\partial s} \right] + p'(s) J = 0. \label{eq14a}
\end{align}
These equations are expressed in terms of the covariant components of $\BFB$,
\be \label{eq10}
     \BFB = B_s \nabla s + B_u \nabla u + B_v \nabla v,
\ee
where
\be
     B_s = \BFT_s \cdot \BFB, \quad
     B_u = \BFT_u \cdot \BFB, \quad
     B_v = \BFT_v \cdot \BFB.
\ee
We note that the current density $\BFJ = \nabla \times \BFB$ can be written as
\be
  \BFJ = J^s \BFT_s + J^u \BFT_u + J^v \BFT_v,
\ee
where the contravariant components of $\BFJ$ are
\be
    J^s = \frac{1}{J} \left( \frac{\partial B_v}{\partial u} -
                            \frac{\partial B_u}{\partial v} \right), \quad
   J^u = \frac{1}{J} \left( \frac{\partial B_s}{\partial v} -
                             \frac{\partial B_v}{\partial s} \right), \quad  
   J^v = \frac{1}{J} \left( \frac{\partial B_u}{\partial s} -
                             \frac{\partial B_s}{\partial u} \right).        
\ee
%
These expressions give us clear interpretations for \Eqn{13a} and \Eqn{14a}. \Eqn{13a} expresses the fact that the force balance condition $\BFJ \times \BFB = \nabla p$ implies that $J^s = \BFJ \cdot \nabla s = 0$. \Eqn{14a} is the $s$-component of the force balance (expressed
in covariant form).

We denote by $Jp(s,u,v)$ the quantity
\be \label{eq:Jp}
      Jp = \frac{\BFJ \cdot \BFB}{|\BFB|^2} = \frac{J^u B_u + J^v B_v}{|\BFB|^2},
      \quad |\BFB|^2 = B^u B_u + B^v B_v.
\ee
In a slight abuse of vocabulary, for the remainder of the article we will simply refer to $Jp$ 
as the parallel current, although
the actual parallel current density 
has instead a single power of $|\BFB|$ in the denominator.

Following the normalization for $\psi(s,u,v)$ adopted in \cite{Tayl94}, we 
write
\be \label{eq17a}
    \psi(s,u,v) = \pi [u - \iota(s) v] + \tilde{\psi}(s,u,v),
\ee
where $\tilde{\psi}$ is periodic in $u$ and $v$. Although $\psi$ is multi-valued, from \Eqn{4} this  representation leads to a single-valued magnetic field with poloidal and toroidal  fluxes determined by the rotational transform $\iota(s)$. We also note that \Eqn{13a} determines $\tilde{\psi}$
only up to an arbitrary function of $s$, which we specify by requiring the mean Fourier harmonic  $\tilde{\psi}_{00}(s)$ to vanish.
\begin{figure}[h]
\begin{center}
\resizebox{3.5in}{!}{\includegraphics{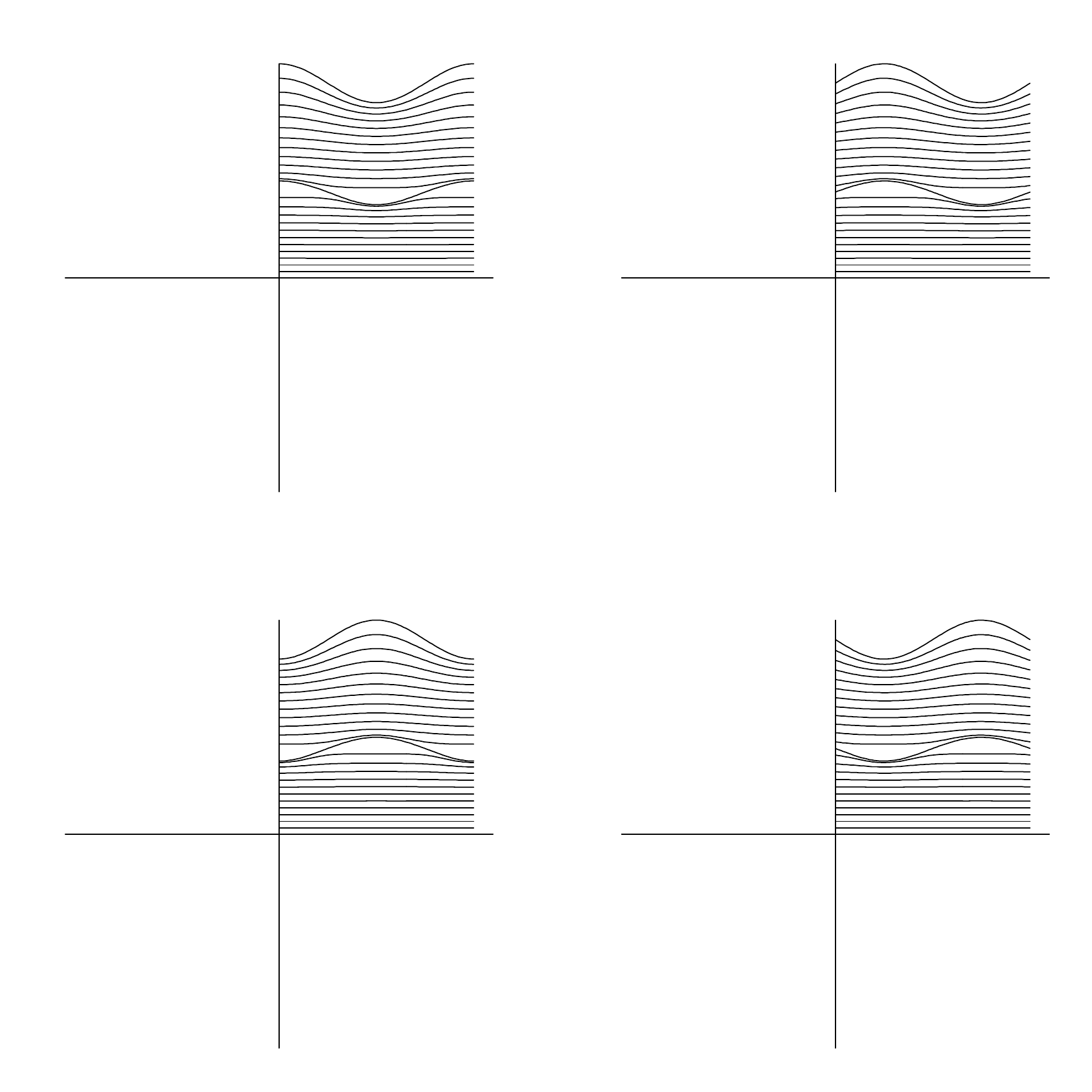}}
\parbox{6.0in}{
\caption{NSLAB computation of a slab equilibrium as described in section \ref{sec:NSTAB}, for $L_{x}=L_{y}=1$, the pressure profile $p(s) = 0.01 (1 - s^2)^2$, the rotational transform profile $\iota(s) = 0.5 + 1.0 s$, a flat lower flux surface, 
$z_0(u,v) = 0$, and an upper flux surface given by $z_1(u,v) = 1 + 0.1 \cos 2 \pi(u - v)$. The plots show various flux surfaces $z = z(s,u,v)$ 
for constant $s$ versus the poloidal angle $u$ at four toroidal angles $v = 0$, $v = 1/4$, $v = 1/2$, and $v = 3/4$ (from left to right and from top to bottom).}
\label{fig00} }
\end{center}
\end{figure}

\subsection{Numerical Scheme}

The nonlinear governing equations (\ref{eq13a}) and (\ref{eq14a}) are solved numerically following the procedure given in \cite{Tayl94}.
In brief, the equations are discretized using second-order-accurate finite differences in $s$, with a 
pseudospectral representation in the angular variables $u$ and $v$. 
A staggered mesh in $s$ is employed, with $L_2(R)$ evaluated at nodes, and $L_1(\psi)$ evaluated at centers; this allows a conservative difference scheme with a compact stencil \cite{BaBG78} that can capture singularities over two or
three mesh points, as illustrated in the numerical results below.

A second-order Richardson method
is used to solve the resulting equations iteratively. This scheme can be viewed as introducing
an artificial time $t$, and solving
\be \label{eq15a} 
     a_\psi \frac{\partial^2 \psi}{\partial t^2} + e_\psi \frac{\partial \psi}{\partial t} = L_1(\psi), \quad
     a_R \frac{\partial^2 R}{\partial t^2} + e_R \frac{\partial R}{\partial t} = L_2(R), 
\ee
via an explicit-in-time discretization with $t = n \Delta t$. The constants $a_\psi$ and $a_R$ are 
chosen to maintain numerical stability of the scheme with the time step $\Delta t$ on the same order as the spatial mesh,
and the coefficients $e_\psi$ and $e_R$ are chosen dynamically to optimize convergence \cite{Beta88}. In practice, the right hand sides of \Eqn{15a} are preconditioned to accelerate convergence, 
and the iteration is actually performed in Fourier space by updating the Fourier coefficients of $R$ and $\psi$ with respect to the angular coordinates \cite{Tayl94}.

An example of a numerical solution with a resonant flux surface computed using NSLAB is shown in Fig.~2, demonstrating that the slab geometry also supports singular behavior similar to that observed using NSTAB.
The spacing of the flux surfaces reflects a resonance where $\iota(s) = 1$,
and the surfaces display a corresponding symmetry with poloidal and
toroidal mode numbers $(m,n) = (1,1)$.
Further examples will be discussed in more detail in Section 5. We first include a discussion of the singularities present in a linerarized treatment of small amplitude perturbations of a planar geometry.

%
%

\section{Linearized Equations}

To illustrate the resonances at rational flux surfaces, it is useful to consider the linearized governing equations
for small-amplitude perturbations of two planar flux surfaces bounding the plasma.
specifically, we consider a perturbation expansion of the MHD equilibrium relative to a one-dimensional
base state corresponding to flat walls $z_0 = a_0$ and $z_1 = b_0$.
We consider a normal mode perturbation of the system with wavenumber $(m,n)$, 
with a small expansion parameter $\epsilon$, $|\epsilon| \ll 1$, which results in a 
linear problem at first order in $\epsilon$. The expansion is performed with the aid of a computer algebra system;
we omit the details and summarize the results.

The perturbed bottom and top walls are assumed to take the form
\be
    z_0(u,v) = a_0 + \epsilon a_{mn} \cos 2 \pi (m u - nv), \quad
    z_1(u,v) = b_0 + \epsilon b_{mn} \cos 2 \pi (m u - n v),
\ee
respectively, and their difference $z_2(u,v) = z_1(u,v) - z_0(u,v)$ is denoted by
\be
    z_2(u,v) = c_0 + \epsilon c_{mn} \cos 2 \pi (m u - n v),
\ee
with $c_{mn} = b_{mn} - a_{mn}$. The corresponding expansion for $R(s,u,v)$ is
\be
    R(s,u,v) = R_0(s) + \epsilon R_{mn}(s) \, \cos 2 \pi (m u - n v) + O(\epsilon^2),
\ee
and that for $\psi(s,u,v)$ is
\be
    \psi(s,u,v) = \pi [u - \iota(s) v] + 
         \epsilon \psi_{mn}(s) \, \sin 2 \pi (m u - nv) + O(\epsilon^2).
\ee
Note the presence of the $\sin$ function as opposed to the $\cos$ function for $R$ for the normal mode representation of $\psi(s,u,v)$, which corresponds to the difference in the number of derivatives appearing for $\psi(s,u,v)$ and $R(s,u,v)$ in the governing equations.

\subsection{Base State}

Expanding in $\epsilon$ gives the leading order nonlinear 
ordinary differential equation for the one dimensional base state, 
\be \label{eq7}
     R_0''(s) - \left[ \frac{L_x^2 \, \iota(s) \, \iota'(s)}{(L_y^2 + 
    [\iota(s)]^2 \, L_x^2)} \right] R_0'(s) -
      \left[\frac{c_0^2 L_x^2 \, L_y^2 \, p'(s)}{\pi^2 (L_y^2 + 
    [\iota(s)]^2 \, L_x^2)}\right] [R_0'(s)]^3 = 0,
\ee
with $R_0(0) = 0$ and $R_0(1) = 1$. The solution depends on the dimensions $a_{0}$, $b_{0}$, $L_{x}$ and $L_{y}$ of the system, the rotational transform $\iota(s)$, and the
pressure gradient $p'(s)$. In the force-free case with $p'(s) = 0$, and
with zero shear, $\iota'(s) = 0$, the solution is just $R_0(s) = s$. The general
case requires the numerical solution of this nonlinear differential equation.

\subsection{First Order Equations}

At first order, we obtain a linear equation that can be solved for the perturbation $\psi_{mn}(s)$ in terms of
$c_{mn}/c_0$ and $R_{mn}'(s)/R_0'(s)$,
\be \label{eq15}
   \psi_{mn}(s) =  \frac{[L_y^2 m + L_x^2 n \iota(s)]}{2 [L_y^2 m^2 + L_x^2 n^2]}
     \left[ \frac{c_{mn}}{c_0} + \frac{R_{mn}'(s)}{R_0'(s)} \right]. 
\ee
The perturbation $R_{mn}(s)$ satisfies the linear second order ordinary differential equation
\be \label{eq17}
    \alpha_1 R_{mn}''(s) + \alpha_2 R_{mn}'(s) + \alpha_3 R_{mn}(s)  + \gamma  = 0,
\ee
where the coefficients in these equations are
\be
    \alpha_1 = \frac{-L_x L_y \pi^2 [n -  m \iota(s)]^2}
                  {c_0 (L_y^2 m^2 + L_x^2 n^2) [R_0'(s)]^2},
 \ee
\be \label{eq22a}
   \alpha_2 = \frac{L_x L_y \pi^2 [L_x^2 (3 n \iota(s) 
       - m \iota(s)^2) + 2 L_y^2 m] \, [n - m \iota(s)] \, \iota'(s)}
   {c_0 (L_y^2 m^2 + L_x^2 n^2) (L_y^2 + L_x^2 [\iota(s)]^2) [R_0'(s)]^2}
\ee
\[
   \mbox{} + \frac{3 c_0 L_x^3 L_y^3 [n - m \iota(s)]^2 p'(s)}
   {(L_y^2 m^2 + L_x^2 n^2) (L_y^2 + L_x^2 [\iota(s)]^2)}
\]
\be
   \alpha_3 = \frac{4 c_0 \pi^4 [n - m \iota(s)]^2)}{L_x L_y},
\ee
and the inhomogeneous term is
\be \label{eq21}
    \gamma= \frac{2 c_{mn} L_x^3 L_y^3 \, R_0'(s) \, [n - m \iota(s)]^2 \, p'(s)}
    {(L_y^2 m^2 + L_x^2 n^2) (L_y^2 + L_x^2 [\iota(s)]^2)} 
\ee
\[
 \mbox{} + \frac{2 c_{mn} L_x L_y \pi^2 
    \, [L_y^2 m + L_x^2 n] \, [n - m \iota(s)] \, \iota(s) \, \iota'(s)}
    {c_0^2 (L_y^2 m^2 + L_x^2 n^2) (L_y^2 + L_x^2 [\iota(s)]^2) R_0'(s)} 
\]
\[ \mbox{} + \frac{4 \pi^4 \, [a_{mn} + c_{mn} R_0(s)] \, [n - m \iota(s)]^2}{L_x L_y}.
\] 
Note the common appearance of the resonant factor $[n - m \iota(s)]$ in each coefficient. In particular, the
terms involve the factors $[n - m \iota(s)] \iota'(s)$ 
and $[n - m \iota(s)]^2 p'(s)$. For vanishing 
shear, $\iota'(s) = 0$, each remaining 
term contains a quadratic factor of $[n - m \iota(s)]^2$, and the singularity at
$\iota(s) = n/m$ is removable. This result is in agreement with the well-known result that nonsymmetric equilibria with nested flux surfaces can be constructed for constant rotational transform \cite{Weitzner14,Lortz1970}. On the other hand, with
moderate shear $\iota'(s_0) \ne 0$ at the resonant surface
$s = s_0$, \Eqn{17} is singular at $s_0$, 
and the leading order behavior of the singularity does not change qualitatively with changes in $p'(s_0)$. 
The case of small shear is thus a
singular limit of \Eqn{17}, and
finite pressure effects can be significant in this case.

\subsection{Numerical Example}
\label{sec:linearized_example}
\begin{figure}[t]
\begin{center}
\resizebox{4.0in}{!}{\includegraphics{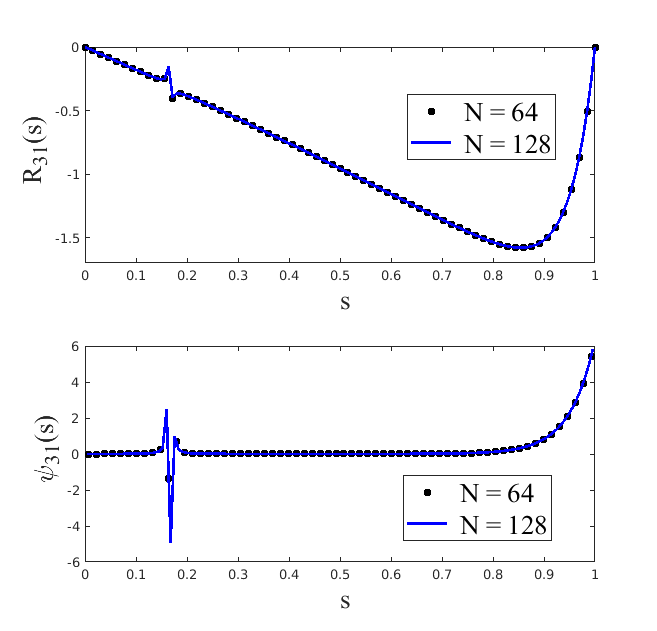}}
\parbox{5.5in}{
\caption{Numerical solution of Eqs.~(\ref{eq15}) and (\ref{eq17})
for $R_{31}(s)$ and $\psi_{31}(s)$ for the parameters and profiles given in Section \ref{sec:linearized_example} using 64 grid points (black dots) and 128 grid points
(blue curves). A singularity is present at the surface where $\iota(s) = 1/3$.}
\label{fig3} }
\end{center}
\end{figure}
As an example, we take $L_x = L_y = 1$, $a_0 = a_{mn} = 0$, $b_0 = b_{mn} = 1$, and
\be
     \iota(s) = 0.25 + 0.5 \, s, \quad p(s) = 1.5 \, s (1 - s),
\ee
and we consider the resonant surface where $\iota(s) = 1/3$ by adding a $(3,1)$ harmonic to the external boundary. 
A finite difference solution for $R_{31}(s)$, and the corresponding solution $\psi_{31}(s)$,
is shown in Figure 3. We do not attempt any special treatment of the singularity in this simple case, 
since this is consistent with the specific finite difference scheme employed in NSTAB and NSLAB, and 
the solution in Fig.~3 should therefore reproduce the behavior expected in those codes for
small amplitude perturbations. In this case, the resonant surface where $\iota(s) = 1/3$ at $s \approx 0.167$
does not lie on the numerical grid, and the solution is exhibiting singular behavior at nearby mesh points.
Note that there are continuous gradients in $R_{31}(s)$ (and in $\psi_{31}(s)$, which is coupled to $R_{31}'(s)$) near the walls where 
$R_{31}(s)$ vanishes,
although these smooth variations are easily distinguished from the singular behavior at the resonant surface.

\section{NSLAB Numerical results}

We start the discussion of the NSLAB numerical results 
with a comparison of the linearized results from the previous section with a corresponding nonlinear NSLAB computation
for a small-amplitude, single-mode perturbation of the upper wall.
Our examples will all feature a flat lower surface, $z_0(u,v) = 0$, with
$L_x = L_y = 1$.

\subsection{Comparison of NSLAB and Linearized Results}\label{sec:comparison}

We consider a force-free case with $p'(s) = 0$ and rotational transform
$\iota(s) = 0.95 + 0.1 s$, focusing on a $(1,1)$ mode at the resonant surface $s_0 = 1/2$. 
\begin{figure}[h]
\begin{center}
\resizebox{4.0in}{!}{\includegraphics{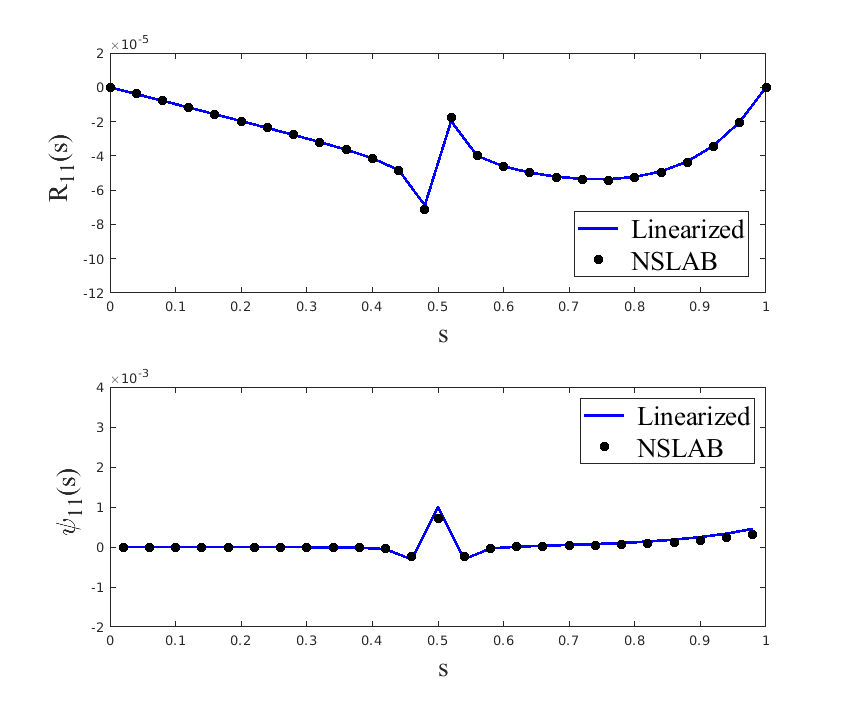}}
\parbox{6.0in}{
\caption{Comparison between the linearized solution (blue curve) and the corresponding NSLAB computation (black dots) for the slab equilibrium described in Section \ref{sec:comparison}.}
\label{fig4} }
\end{center}
\end{figure}
The  upper surface is given by
\be    
     z_1(u,v) = 1 + \Delta_{11} \cos 2 \pi (u - v), 
\ee
with amplitude $\Delta_{11} = 10^{-4}$. The comparison is
illustrated in Fig. 4, where we show $\psi_{11}$ given by the analytic linear calculation and
the $(1,1)$ Fourier harmonic of NSLAB's nonlinear solution for $\psi(s,u,v)$. On a mesh  of 25 points, the agreement is seen to be quite
satisfactory. The numerical solution for $\psi_{11}(s)$ shows a localized peak at the three mesh points centered at $s_0 = 1/2$.
As a rough indicator of the strength of the singularity, we use 
\be \label{eq29}
      D^2 \psi_{11} = \frac{\psi_{11}(s_0) - [\psi_{11}(s_0 + h) + \psi_{11}(s_0 - h)]/2]}{h^2} 
       \approx \frac{1}{2} \frac{d^2 \psi_{11}}{d s^2}(s_0)
\ee
where $h$ is the mesh spacing in $s$. As well as approximating $\psi_{11}''(s_0)/2$, $D^2 \psi_{11}$ characterizes
the peak amplitude relative to the average value of the two neighboring values. Since $\BFJ = \nabla \times [\nabla \times (\psi \nabla s)]$,
$\psi_{11}''(s_0)$ is an effective measure of the singular current strength. For the case shown in Fig.~4, 
we find $D^2 \psi_{11} = 0.5938$.
At this point, we stress that we recognize that the computation of equilibria with a localized singularity on a fixed mesh is necessarily plagued by relatively high levels
of truncation error. However, we will not be concerned by this numerical issue since our goal for the remainder of the article is to find appropriate wall shapes that, as much as possible, eliminate these singularities. The resulting
smooth solution can then achieve the level of accuracy that is expected of a second-order-accurate finite difference scheme.

\subsection{Eliminating Current Sheets by Wall Modification}

The remainder of the article focuses on our central motivation for this work, namely the elimination current sheets by suitable modifications of the shape of the upper wall. We consider various cases with one, two, and three resonant surfaces for force-free equilibria with $p'(s) = 0$ or for finite pressure equilibria with $p(s) = p_0 (1 - s^2)^2$.
In this study, we consider relatively weak current sheets that can be eliminated by small amplitude perturbations of a flat upper wall. We generally represent the upper wall as a finite Fourier series
\be
       z_1(u,v) = \sum_{m,n} \Delta_{mn} \cos 2 \pi (m u - n v)
\ee
where the mean position of the wall is $\Delta_{00} = 1$. The magnitude of the pressure in the finite-pressure equilibria we will study will be expressed in terms of the usual $\beta$ parameter, defined by
\be
    \beta = \frac{2 \int_V p \, dV}{\int_V |\BFB|^2 dV},
\ee
and which can be determined from a numerical integration once the
solution has been computed.

\subsubsection{Single Resonant Surface}
\begin{figure}[ht]
\begin{center}
\resizebox{5.0in}{!}{\includegraphics{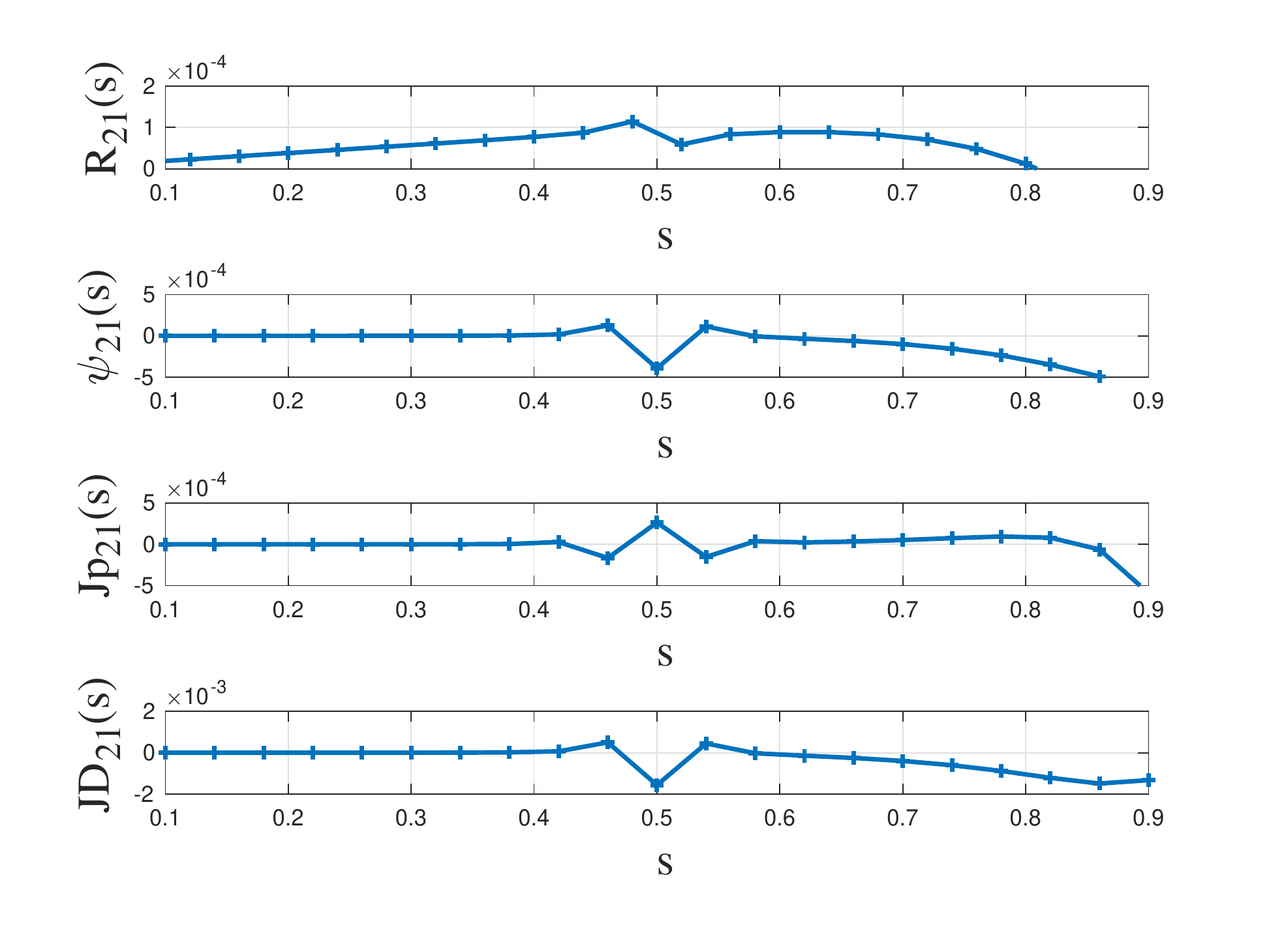}}
\parbox{6.0in}{
\caption{Slab equilibrium with a single resonant surface. The rotational transform profile is $\iota(s)= 0.35+0.3 s$, the pressure profile is $p(s)= 0.3 (1 - s^2)^2$, $\beta = 2.5\%$, and the wall perturbations correspond to
$\Delta_{20} = \Delta_{01}=0.01$ and $\Delta_{21}=0$.
From top to bottom, we plot the Fourier component $R_{21}(s)$ of $R(s,u,v)$,
the Fourier component $\psi_{21}(s)$ of $\psi(s,u,v)$,
the Fourier component $Jp_{21}(s)$ of the parallel current $Jp(s,u,v)$,
and the Fourier component $JD_{21}(s)$ of the Jacobian $J(s,u,v)$, all as a 
function of flux surface coordinate $s$.}
\label{fig:Jp1} }
\end{center}
\end{figure}

For the case of a single resonant surface, we consider 
the rotational transform profile $\iota(s) = 0.35 + 0.3 s$. To generate a (2,1) current sheet, we start with a perturbed upper wall with $(2,0)$ and $(0,1)$ harmonics,
\be   
    z_1(u,v) = 1 + 0.01 \cos 4 \pi u + 0.01 \cos 2 \pi v.
\ee
The nonlinear interaction of the $(2,0)$ and $(0,1)$ modes 
is found to generate at quadratic order
a $(2,1)$ mode that triggers a current sheet at the resonant surface $s_0 = 1/2$, where $\iota(s_0)=1/2$.

Some numerical results are given in 
\Fig{fig:Jp1} for a case with
$p_0 = 0.3$ corresponding to $\beta = 2.5\%$.
The top two plots in the figure show the $(2,1)$ Fourier components $R_{21}(s)$ and $\psi_{21}(s)$ of the computed solutions $R(s,u,v)$ and $\psi(s,u,v)$.
It is also insightful to consider the profiles for two other quantities which have an immediate physical interpretation, namely the parallel current density 
and the Jacobian of the coordinate transformation, which can be viewed as a measure of the distortion of the flux surfaces associated with the appearance of a current sheet.
The $(2,1)$ Fourier
component of the parallel current, $Jp_{21}(s)$, and that of the
Jacobian, $JD_{21}(s)$, are shown as the bottom two plots in  \Fig{fig:Jp1}.
Both the parallel current and the Jacobian profiles show singular behavior that is qualitatively similar to $\psi_{21}(s)$, with peaked singularities that are 
localized near $s = s_0$. In our computations, we generally find that the behavior of the
parallel current profiles $Jp_{mn}(s)$ is faithfully mirrored by that of 
the $\psi_{mn}(s)$ profiles.

\begin{figure}[ht]
\begin{center}
\resizebox{6.0in}{!}{\includegraphics{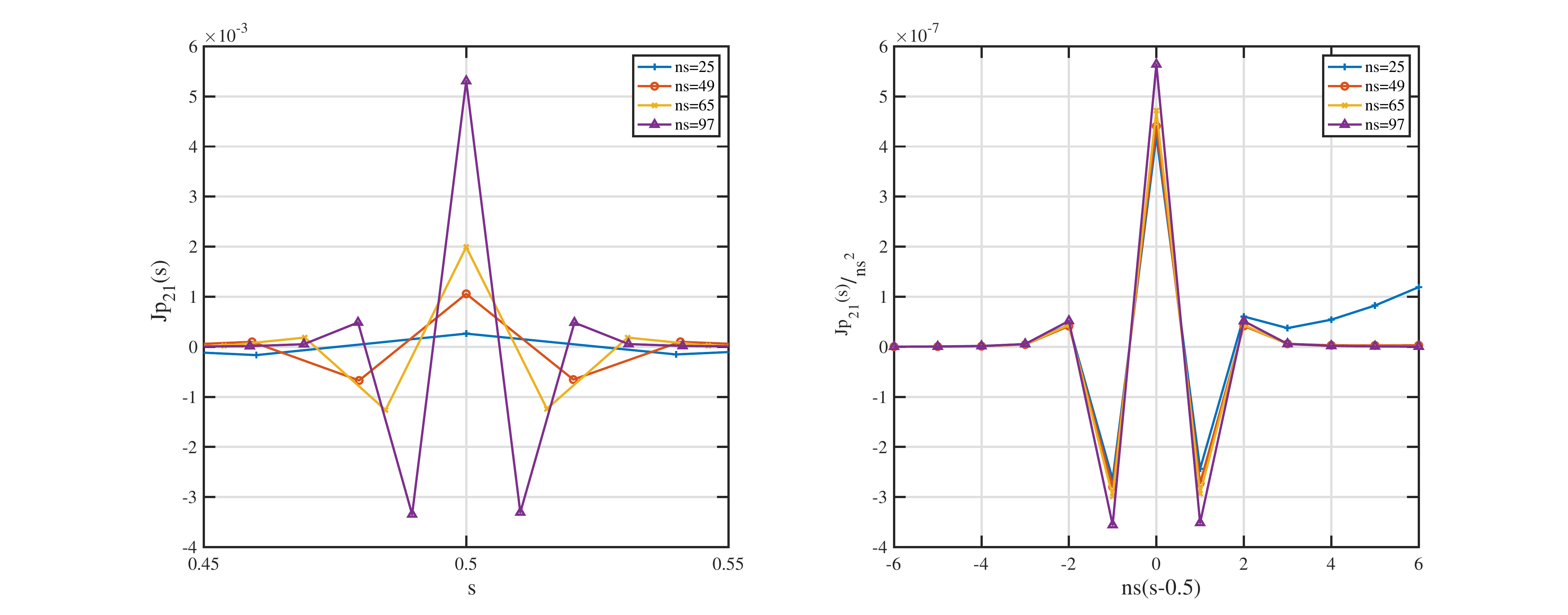}}
\parbox{6.0in}{
\caption{Single resonant surface : the rotational transform $\iota(s)= 0.35+0.3 s$
and the pressure field $p(s)= 0.3 (1 - s^2)^2$, with $\beta = 2.5\%$.
The Fourier component $Jp_{21}(s)$ of the parallel current $Jp(s,u,v)$ (see \eqref{eq:Jp}),
as a function of flux surface $s$ with 
the wall perturbations 
$\Delta_{20} = \Delta_{01}=0.01$ and $\Delta_{21}=0$.
The parallel current profiles $Jp_{21}(s)$ near $s=0.5$ for the mesh refinements using $ns = 25$, 49, 65, and 97 points in the $s$ coordinates (left) 
and their scaled versions (right).
}
\label{fig:Jp2} }
\end{center}
\end{figure}

On the left-hand side of \Fig{fig:Jp2}, we show the parallel current profile in the vicinity of the resonant surface $s = s_0$ 
for a series of mesh refinements using $ns = 25$, 49, 65, and 97 mesh points for the $s$ coordinate.
With decreasing mesh size $1/ns$ the peak increases in magnitude, while the width of the peak decreases. The figure on the right-hand side of \Fig{fig:Jp2} shows a scaled version 
of the figure on the left-hand side of \Fig{fig:Jp2}, where the vertical axis 
is scaled by $ns^2$ and the horizontal axis by $1/ns$. We observe that a satisfactory
calculation of the singularity is captured using a relatively crude mesh. This can be
attributed to the use of a carefully designed conservative difference 
scheme in NSLAB (following that used in NSTAB), which avoids smearing the singularity over 
too many neighboring mesh points.

\begin{figure}[h]
\begin{center}
\resizebox{4.0in}{!}{\includegraphics{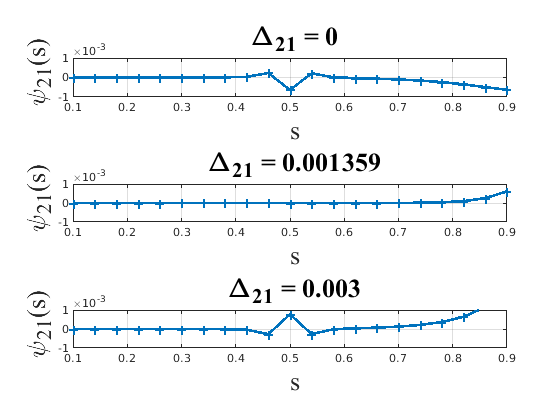}}
\parbox{6.0in}{
\caption{The Fourier component $\psi_{21}(s)$ of $\psi(s,u,v)$ as a 
function of the flux coordinate $s$ for slab equilibria with rotational transform profile $\iota(s)= 0.35+0.3 s$, pressure profile $p(s)= 0.68 (1 - s^2)^2$, so that $\beta = 5.8\%$, wall perturbation amplitudes 
$\Delta_{20} = \Delta_{01}=0.01$, and three different amplitudes for $\Delta_{21}$: $\Delta_{21}=0$ (top),  $\Delta_{21}= 0.000327101$ (middle), and $\Delta_{21}= 0.003$ (bottom).}
\label{fig:pltPSI1_gbm} }
\end{center}
\end{figure}

We next consider the feasability of eliminating the singularity at the $s_{0}=1/2$ surface by varying the fundamental harmonic $\Delta_{21}$ of the wall perturbation.
We consider $p_0 = 0.68$, corresponding to a high-beta equilibrium with $\beta = 5.8\%$.
The profile for the $(2,1)$ Fourier component $\psi_{21}(s)$ of $\psi(s,u,v)$ for $\Delta_{21}=0$ is shown in the top figure in \Fig{fig:pltPSI1_gbm}. The quadratic interaction of the $(2,0)$ and $(0,1)$ wall perturbations has generated a small-amplitude current sheet 
with a negative value of $\psi_{21}(s_0)$. The corresponding value of $D^2 \psi_{21}$ in
\Eqn{29} is $D^2 \psi_{21} = -1.076$.
\begin{figure}[ht]
\begin{center}
\resizebox{4.0in}{!}{\includegraphics{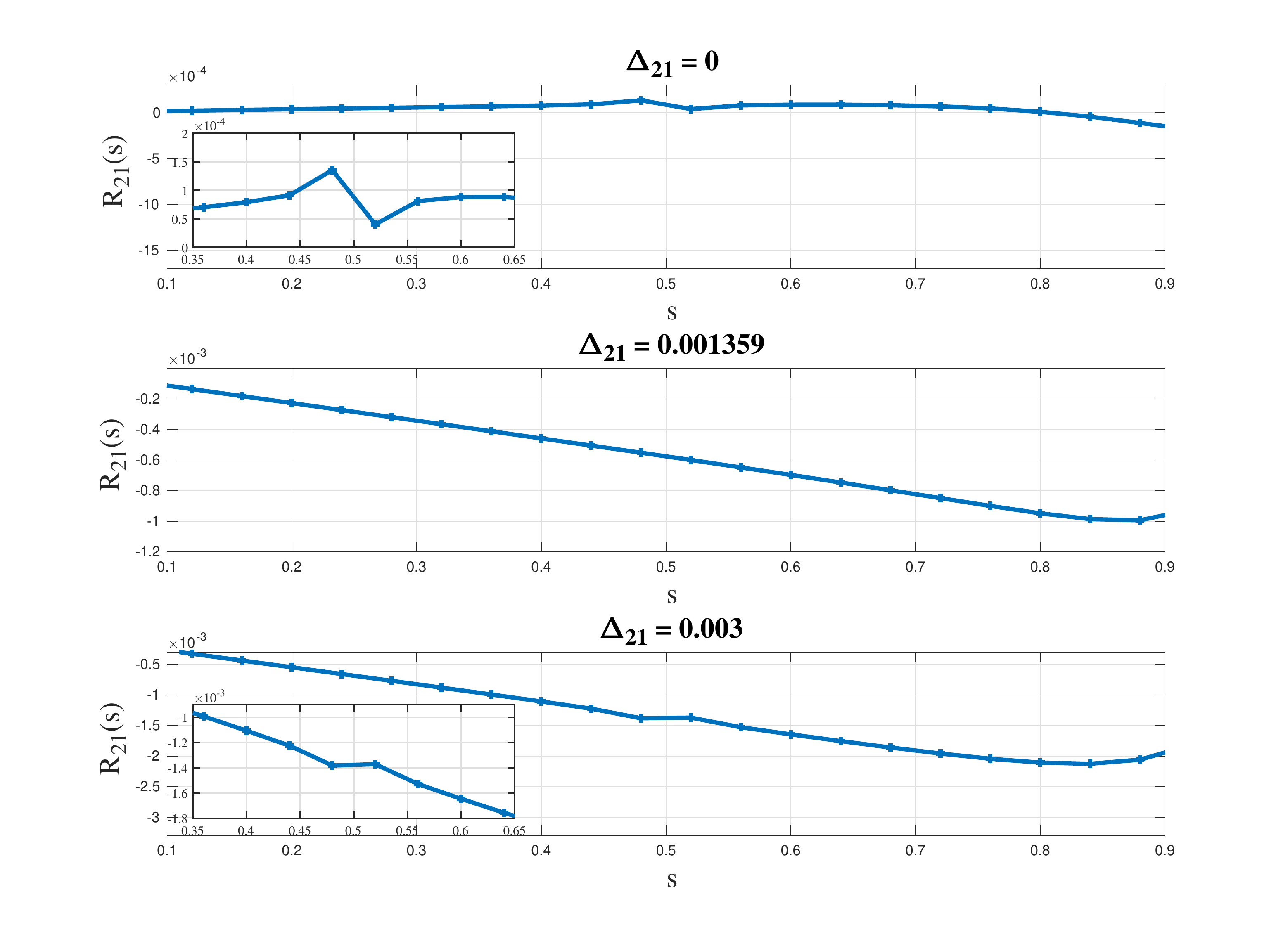}}
\parbox{6.0in}{
\caption{The Fourier component $R_{21}(s)$ of $R(s,u,v)$ as a 
function of the flux coordinate $s$ for slab equilibria with rotational transform profile $\iota(s)= 0.35+0.3 s$, pressure profile $p(s)= 0.68 (1 - s^2)^2$, so that $\beta = 5.8\%$, wall perturbation amplitudes 
$\Delta_{20} = \Delta_{01}=0.01$, and three different amplitudes for $\Delta_{21}$: $\Delta_{21}=0$ (top),  $\Delta_{21}= 0.000327101$ (middle), and $\Delta_{21}= 0.003$ (bottom).}
\label{fig:pltR1_gbm} }
\end{center}
\end{figure}
If we then explicitly introduce a (2,1) wall perturbation of amplitude $\Delta_{21}$, so that
\be   
    z_1(u,v) = 1 + 0.01 \cos 4 \pi u + 0.01 \cos 2 \pi v + \Delta_{21} \cos 2 \pi (2u - v)
\ee
we find that the peak in $\psi_{21}(s_0)$ monotonically 
increases from
negative values, through zero, and then on to positive values as $\Delta_{21}$ is increased 
from zero through positive values. For example, the bottom-most plot in \Fig{fig:pltPSI1_gbm}
for $\Delta_{21} =  0.003$ shows a positive peak in $\psi_{21}(s_0)$, with
$D^2 \psi_{21} = 1.301$. At the intermediate value $\Delta_{21} = 0.00135870$, we
find that $D^2 \psi_{21}$ passes through zero, and the middle figure in \Fig{fig:pltPSI1_gbm} corresponding to that case 
shows a smooth profile for
$\psi_{21}(s)$ in the vicinity of the resonant surface: the singularity has been removed. In \Fig{fig:pltR1_gbm}, we show the corresponding plots for
the $(2,1)$ Fourier component $R_{21}(s)$ of the other dependent variable $R(s,u,v)$, which
also exhibits singular behavior at $s_0$ that is similarly eliminated by modifying the shape of the upper wall.

The critical value of the $(2,1)$ wall perturbation 
$\Delta_{21}$ that eliminates the current sheet with $D^2 \psi_{21} = 0$ can be 
computed by performing a series of NSLAB runs, effectively conducting a root-finding 
search by considering $D^2 \psi_{21}$ to be a function of $\Delta_{21}$. 
More efficiently, this search can instead be incorporated into the overall
NSLAB iterative procedure in \Eqn{15a} by appending an additional evolution equation
\be \label{eq31}
      a_{21}  \frac{d \Delta_{21}}{d t} = - D^2 \psi_{21}
\ee
\begin{figure}[ht]
\begin{center}
\resizebox{4.0in}{!}{\includegraphics{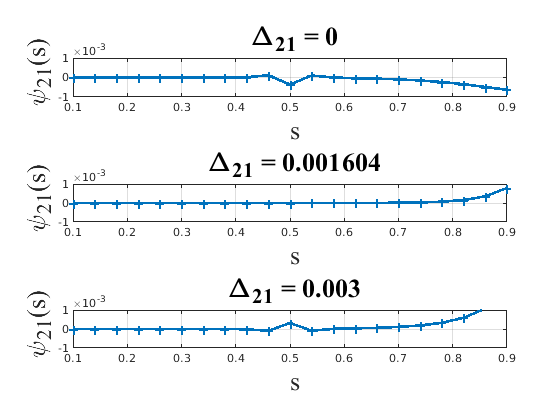}}
\parbox{6.0in}{

\caption{The Fourier component $\psi_{21}(s)$ of $\psi(s,u,v)$ as a 
function of the flux coordinate $s$ for force-free equilibria with rotational transform profile $\iota(s)= 0.35+0.3 s$, wall perturbation amplitudes 
$\Delta_{20} = \Delta_{01}=0.01$, and three different amplitudes for $\Delta_{21}$: $\Delta_{21}=0$ (top),  $\Delta_{21}= 0.00160445$ (middle), and $\Delta_{21}= 0.003$ (bottom).}
\label{fig:pltPSI2_gbm} }
\end{center}
\end{figure}
where $a_{21}$ is a positive relaxation coefficient. In this way, the critical value of
the wall perturbation $\Delta_{21}$ that drives the singularity amplitude $D^2 \psi_{21}$ to zero as the iteration converges can be found in a single NSLAB run.  
\begin{figure}[ht]
\begin{center}
\resizebox{4.0in}{!}{\includegraphics{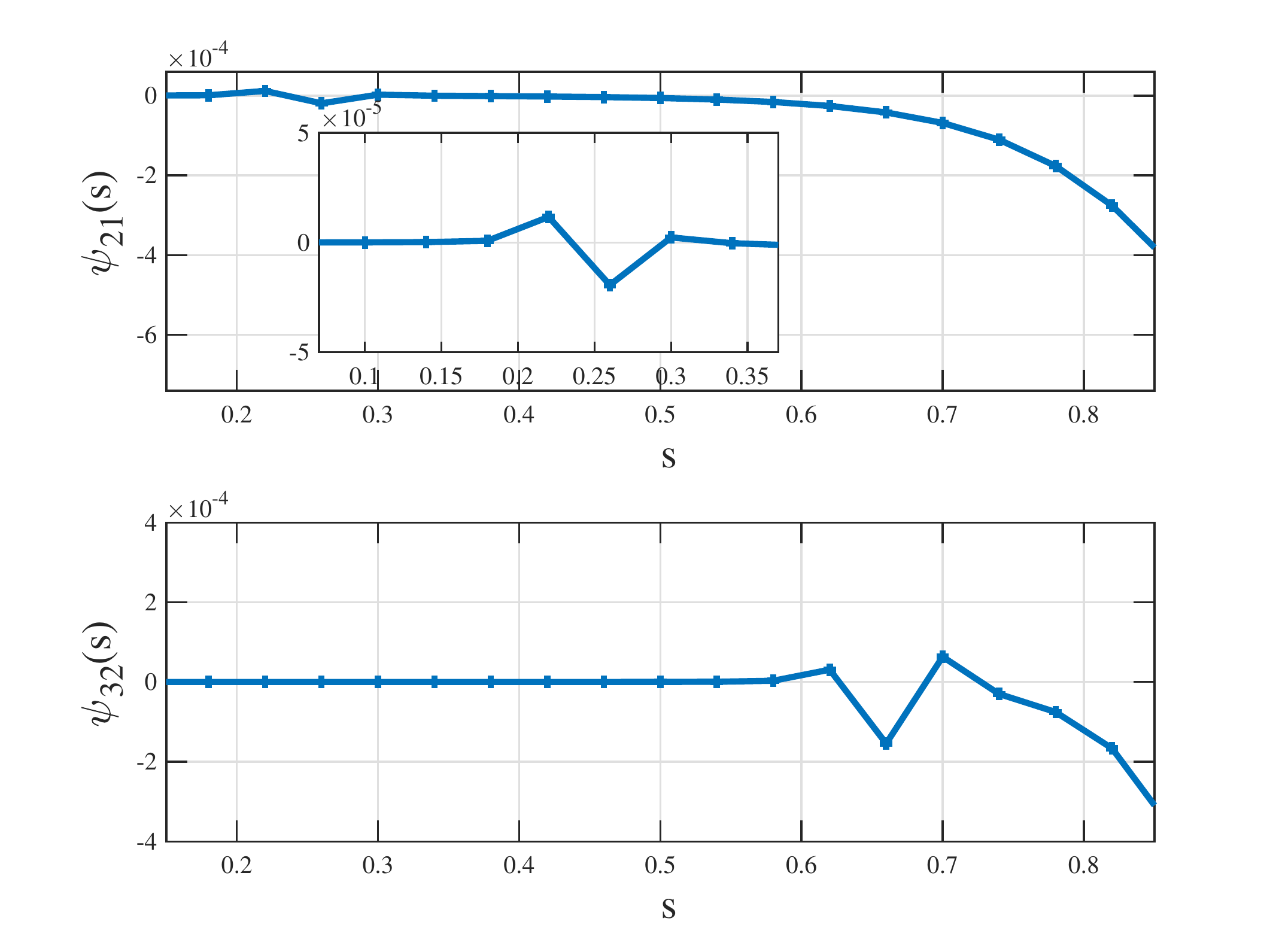}}
\parbox{6.0in}{
\caption{The (2,1) Fourier component $\psi_{21}(s)$ (top) and the (3,2) Fourier component $\psi_{32}(s)$ (bottom) of $\psi(s,u,v)$ as a function of the flux coordinate $s$ for a slab equilibrium with rotational transform profile $\iota(s) = 0.4+0.4s$, the pressure profile $p(s)=0.75(1-s^2)^2$, giving $\beta= 5.8\%$, and wall perturbation amplitudes 
$\Delta_{20}=\Delta_{30}=\Delta_{01}=\Delta_{02}=0.01$ and $\Delta_{21} = \Delta_{32} = 0 $.
See \Fig{fig:two_p1} for the corresponding parallel current $Jp_{mn}$ profiles.
\label{fig:two1}} }
\end{center}
\end{figure}

In \Fig{fig:pltPSI2_gbm}, we again consider the rotational transform profile $\iota(s)=0.35+0.3 s$, for a force-free equilibrium $p'(s)=0$.
In this case, we find an optimal value 
of $\Delta_{21}=0.00160445$ that results in
$D^2 \psi_{21} \approx 0$. Comparing with the corresponding profiles in \Fig{fig:pltPSI1_gbm},
we find that decreasing $\beta$ to zero amplitude has reduced the peaks in $\psi_{21}$
by roughly half; the same is true for the corresponding values of $D^2 \psi_{21}$. We generally
find that the sensitivity of the amplitude of the computed current sheets to $\beta$ is increased
as the shear $\iota'(s)$ decreases. For larger shear, the results tend to become insensitive
to $\beta$; this is consistent with the findings for the linear analysis of normal modes described in the previous section.

\subsubsection{Two Resonant Surfaces}

\begin{figure}[h]
\begin{center}
\resizebox{4.0in}{!}{\includegraphics{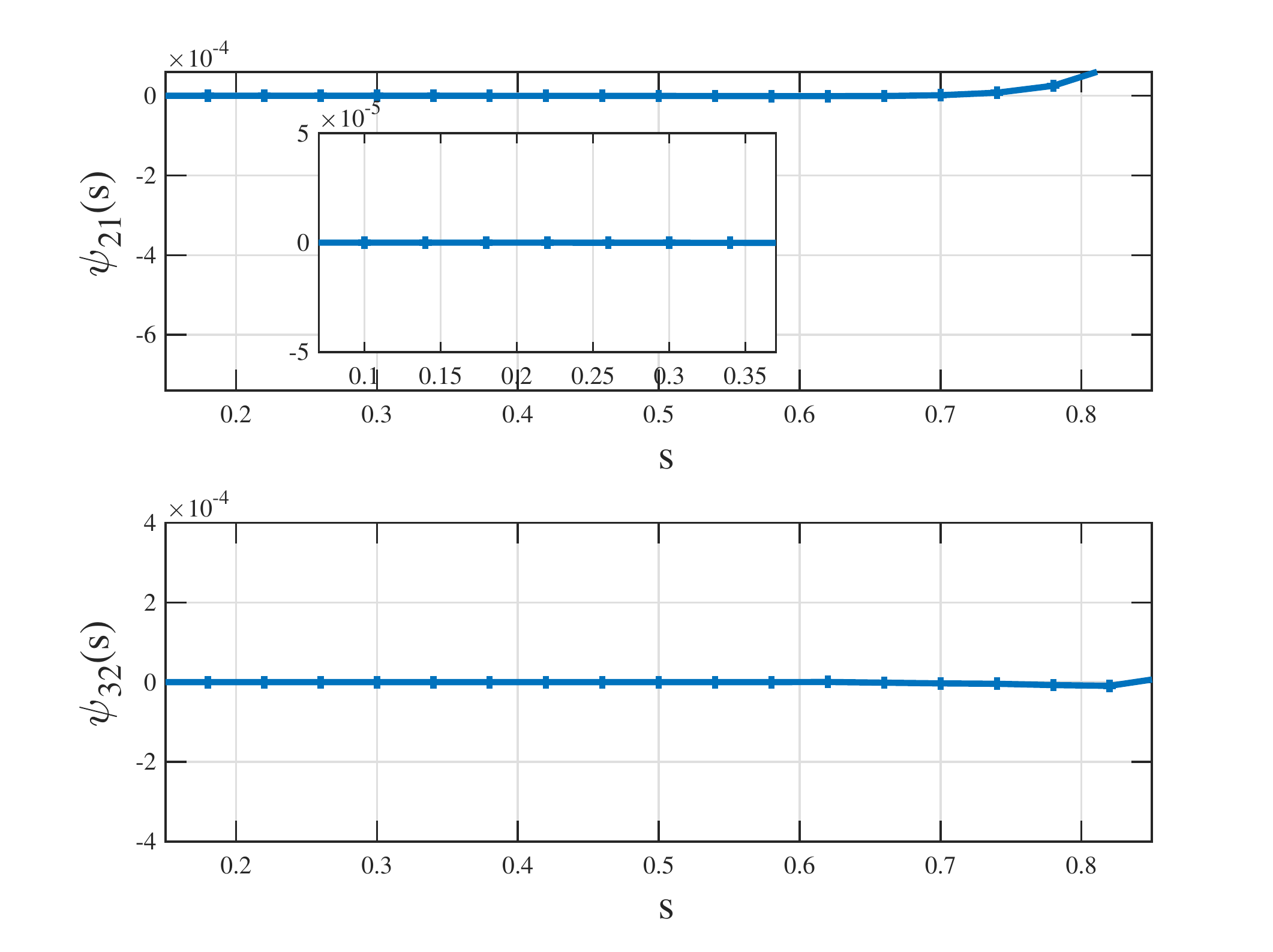}}
\parbox{6.0in}{
\caption{The (2,1) Fourier component $\psi_{21}(s)$ (top) and the (3,2) Fourier component $\psi_{32}(s)$ (bottom) of $\psi(s,u,v)$ as a function of the flux coordinate $s$ for a slab equilibrium with rotational transform profile $\iota(s) = 0.4+0.4s$, pressure profile $p(s)=0.75(1-s^2)^2$, giving $\beta= 5.8\%$, wall perturbation amplitudes 
$\Delta_{20}=\Delta_{30}=\Delta_{01}=\Delta_{02}=0.01$, $\Delta_{21} =  0.001147$, and $\Delta_{32} =  0.001998$. Comparing this figure with figure \ref{fig:two1}, we observe that the singularity has been eliminated.
See \Fig{fig:two_p2} for the corresponding parallel current $Jp_{mn}$ profiles.
\label{fig:two2}}
}
\end{center}
\end{figure}

\begin{figure}[h]
\begin{center}
\resizebox{4.0in}{!}{\includegraphics{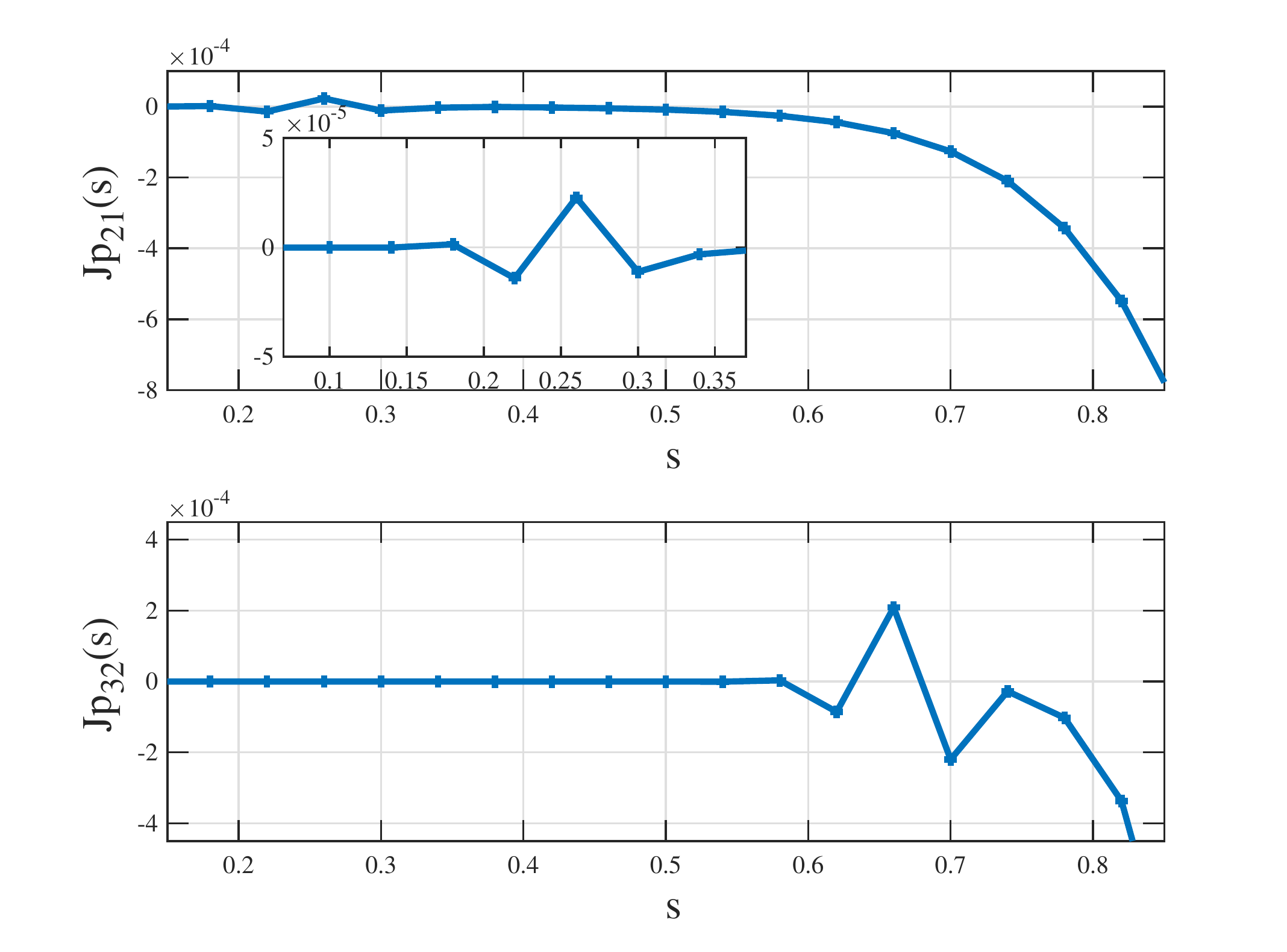}}
\parbox{6.0in}{
\caption{The (2,1) Fourier component $\mathrm{Jp}_{21}(s)$ (top) and the (3,2) Fourier component $\mathrm{Jp}_{32}(s)$ (bottom) of $\mathrm{Jp}(s,u,v)$ as a function of the flux coordinate $s$ for a slab equilibrium with rotational transform profile $\iota(s) = 0.4+0.4s$, and pressure profile $p(s)=0.75(1-s^2)^2$,  corresponding to $\beta= 5.8\%$, and wall perturbation amplitudes 
$\Delta_{20}=\Delta_{30}=\Delta_{01}=\Delta_{02}=0.01$ and $\Delta_{21} = \Delta_{32} = 0 $. 
\label{fig:two_p1}}
}
\end{center}
\end{figure}

\begin{figure}[h]
\begin{center}
\resizebox{4.0in}{!}{\includegraphics{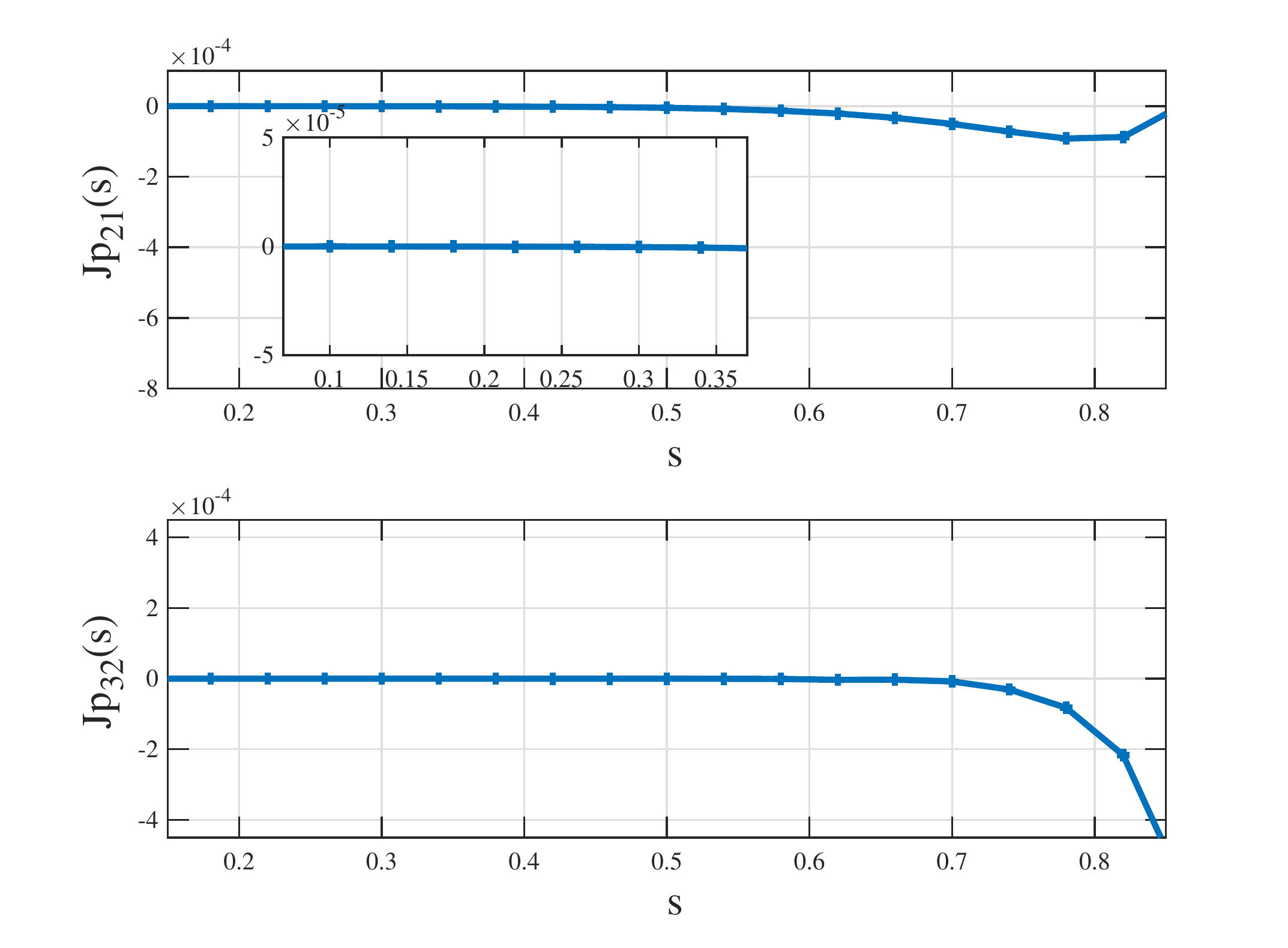}}
\parbox{6.0in}{
\caption{The (2,1) Fourier component $\mathrm{Jp}_{21}(s)$ (top) and the (3,2) Fourier component $\mathrm{Jp}_{32}(s)$ (bottom) of $\mathrm{Jp}(s,u,v)$ as a function of the flux coordinate $s$ for a slab equilibrium with rotational transform profile $\iota(s) = 0.4+0.4s$, and pressure profile $p(s)=0.75(1-s^2)^2$,  corresponding to $\beta= 5.8\%$, and wall perturbation amplitudes $\Delta_{20}=\Delta_{30}=\Delta_{01}=\Delta_{02}=0.01$, $\Delta_{21} =  0.001147$, and $\Delta_{32} =  0.001998$. Comparing these figures with Figure \ref{fig:two_p1}, we see that the current singularity has been eliminated.
\label{fig:two_p2}} }
\end{center}
\end{figure}

We next consider the rotational transform profile $\iota(s)=0.4+0.4s$, which
includes the low-order rationals $\iota = 1/2$ and $\iota = 2/3$, and we trigger
singularities by prescribing the fixed wall perturbations with $(2,0)$ and $(0,1)$ components to generate a $(2,1)$ mode via nonlinear coupling, and wall perturbations with $(3,0)$ and $(0,2)$
components to generate a $(3,2)$ mode, so that we have two prominent resonant surfaces for $\iota=1/2$ and $\iota=2/3$. Specifically, the upper surface is
\be  
   z_1(u,v) = 1 + 0.01 \cos 4 \pi u + 0.01 \cos 2 \pi v + 
   \Delta_{21} \cos 2 \pi (2u - v) 
\ee
\[
     \mbox{} +  0.01 \cos 6 \pi u + 0.01 \cos 4 \pi v + 
     \Delta_{32} \cos 2 \pi (3 u - 2 v).
\]
The resulting equilibrium with pressure profile $p(s) = 0.75(1-s^2)^2$, corresponding to $\beta=5.8\%$, and $\Delta_{21} = \Delta_{32} = 0$ is shown in
\Fig{fig:two1}. 
The profiles for $\psi_{21}(s)$ and $\psi_{32}(s)$ are plotted on similar scales but 
with an inset for $\psi_{21}(s)$ to better show the $(2,1)$ singularity around $s_{21} \approx 0.26$. 
To eliminate the singularities, we generalize the iteration in \Eqn{31} to 
\be \label{eq33}
      a_{21} \frac{d \Delta_{21}}{d t} = - D^2 \psi_{21}, \quad
      a_{32} \frac{d \Delta_{32}}{d t} = - D^2 \psi_{32},
\ee   
where $D^2 \psi_{21}$ and $D^2 \psi_{32}$ are based at $s_{21}$ and $s_{32}$, respectively.
The iteration produces critical wall perturbation values $\Delta_{21} =  0.001147$ and $\Delta_{32} = 0.001998$  
that eliminate the singularities as shown in \Fig{fig:two2}. We mention here that to compute the critical wall perturbation values $\Delta_{21}$ and $\Delta_{32}$, we have also used a quasi-Newton method in a separate run, described in section \ref{sec:three}, and obtained the same results.
We note that the same scales are used in the plots of \Fig{fig:two1} and \Fig{fig:two2}. 
In \Fig{fig:two_p1} and \Fig{fig:two_p2}, we show the profiles of the parallel currents corresponding to \Fig{fig:two1} and \Fig{fig:two2}, respectively. We again observe that the Fourier harmonics $\psi_{21}$ and $\psi_{32}$ of $\psi$, and $Jp_{21}$ and $Jp_{32}$ of $Jp$ have the same behavior, and that we indeed eliminated the current singularity. 

\subsubsection{Three Resonant Surfaces}
\label{sec:three}
\begin{figure}[ht]
\begin{center}
\resizebox{4.0in}{!}{\includegraphics{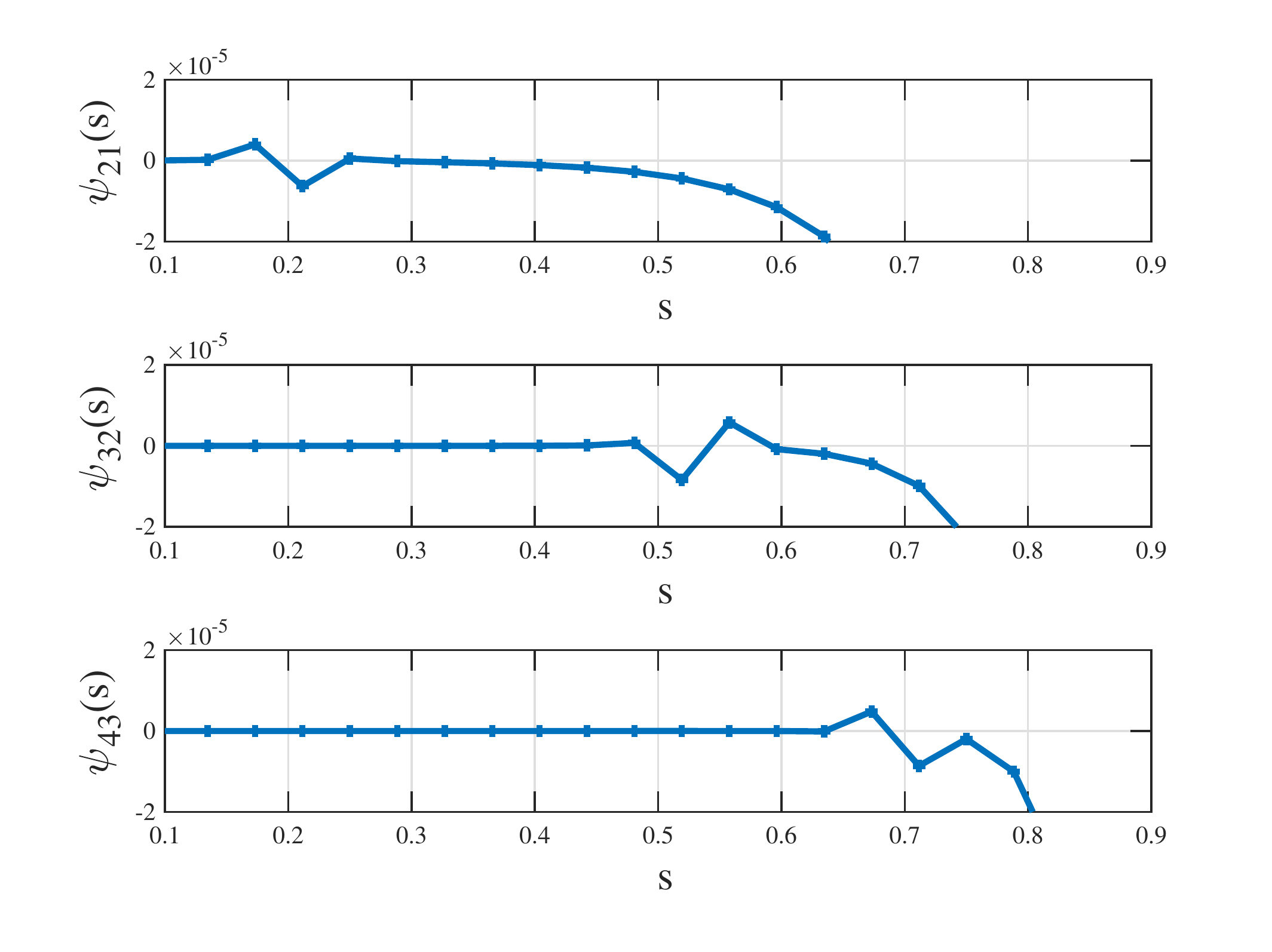}}
\parbox{6.0in}{
\caption{The (2,1) Fourier component $\psi_{21}(s)$ (top), the (3,2) Fourier component $\psi_{32}(s)$ (middle) and the (4,3) Fourier component $\psi_{43}(s)$ (bottom) of $\psi(s,u,v)$ as a function of the flux coordinate $s$ for a force-free slab equilibrium with three resonant surfaces. The rotational transform profile is $\iota(s) = 0.4+0.5s$, and the wall perturbation amplitudes are $\Delta_{20} = \Delta_{30} = \Delta_{40} = \Delta_{01} = \Delta_{02} = \Delta_{03} = 0.01$ and $\Delta_{21} = \Delta_{32} = \Delta_{43} = 0$. See \Fig{fig:threeJ1} for the corresponding parallel current $Jp_{mn}$ profiles.
\label{fig:three1}} }
\end{center}
\end{figure}

\begin{figure}[ht]
\begin{center}
\resizebox{4.0in}{!}{\includegraphics{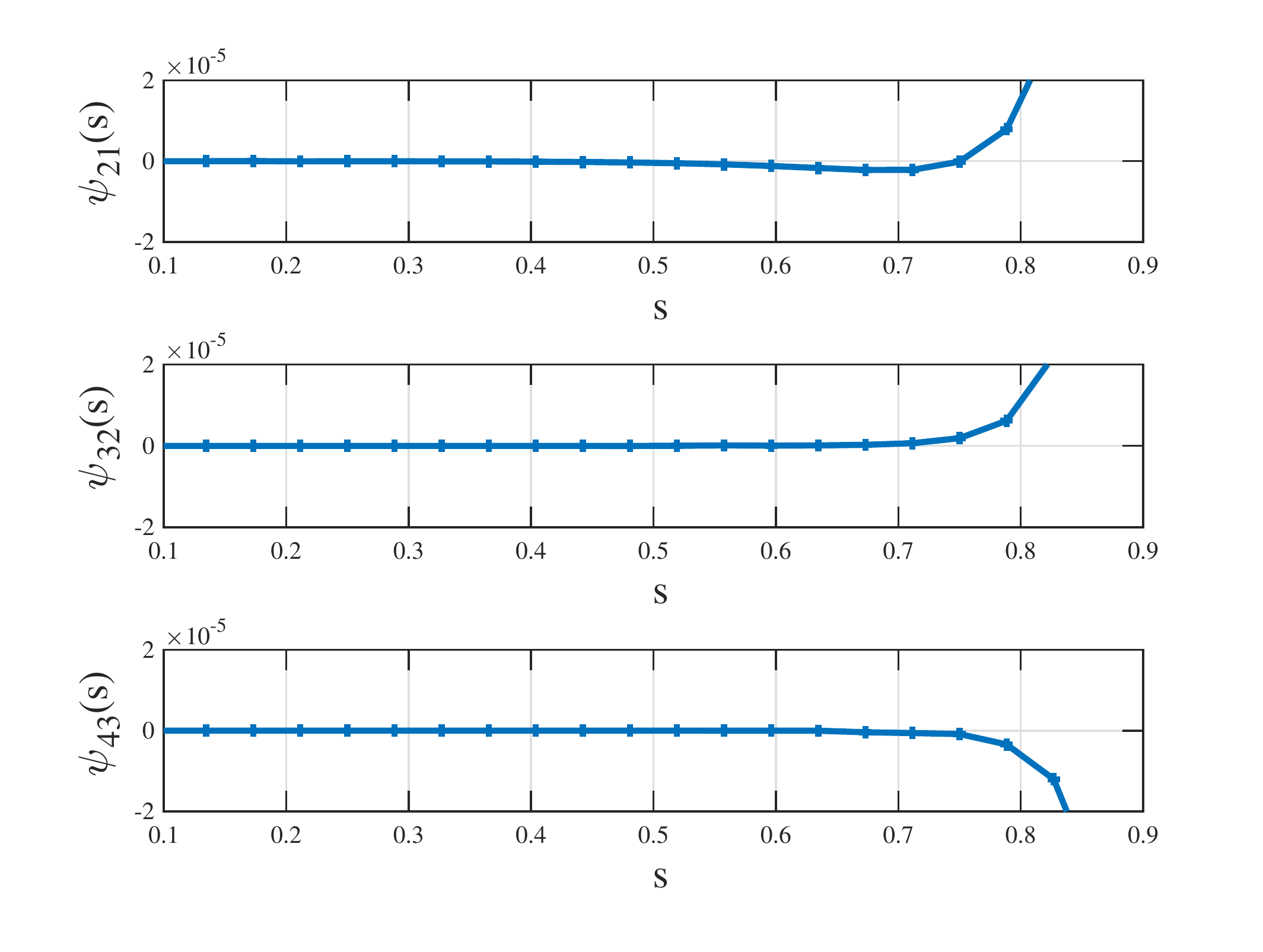}}
\parbox{6.0in}{
\caption{The (2,1) Fourier component $\psi_{21}(s)$ (top), the (3,2) Fourier component $\psi_{32}(s)$ (middle) and the (4,3) Fourier component $\psi_{43}(s)$ (bottom) of $\psi(s,u,v)$ as a function of the flux coordinate $s$ for a force-free slab equilibrium with three resonant surfaces. The rotational transform profile is $\iota(s) = 0.4+0.5s$, and the wall perturbation amplitudes are $\Delta_{20} = \Delta_{30} = \Delta_{40} = \Delta_{01} = \Delta_{02} = \Delta_{03} = 0.01$, $\Delta_{21}=0.00083$, $\Delta_{32}=0.00222$, and $\Delta_{43}=0.00192$. Comparing these figures with Figure \ref{fig:three1}, we see that this particular choice of $\Delta_{21}$, $\Delta_{32}$, and $\Delta_{43}=0.00192$ allowed us to eliminate the singularities at the resonant surfaces. See \Fig{fig:threeJ2} for corresponding parallel current $Jp_{mn}$ profiles.
\label{fig:three2}} }
\end{center}
\end{figure}

\begin{figure}[ht]
\begin{center}
\resizebox{4.0in}{!}{\includegraphics{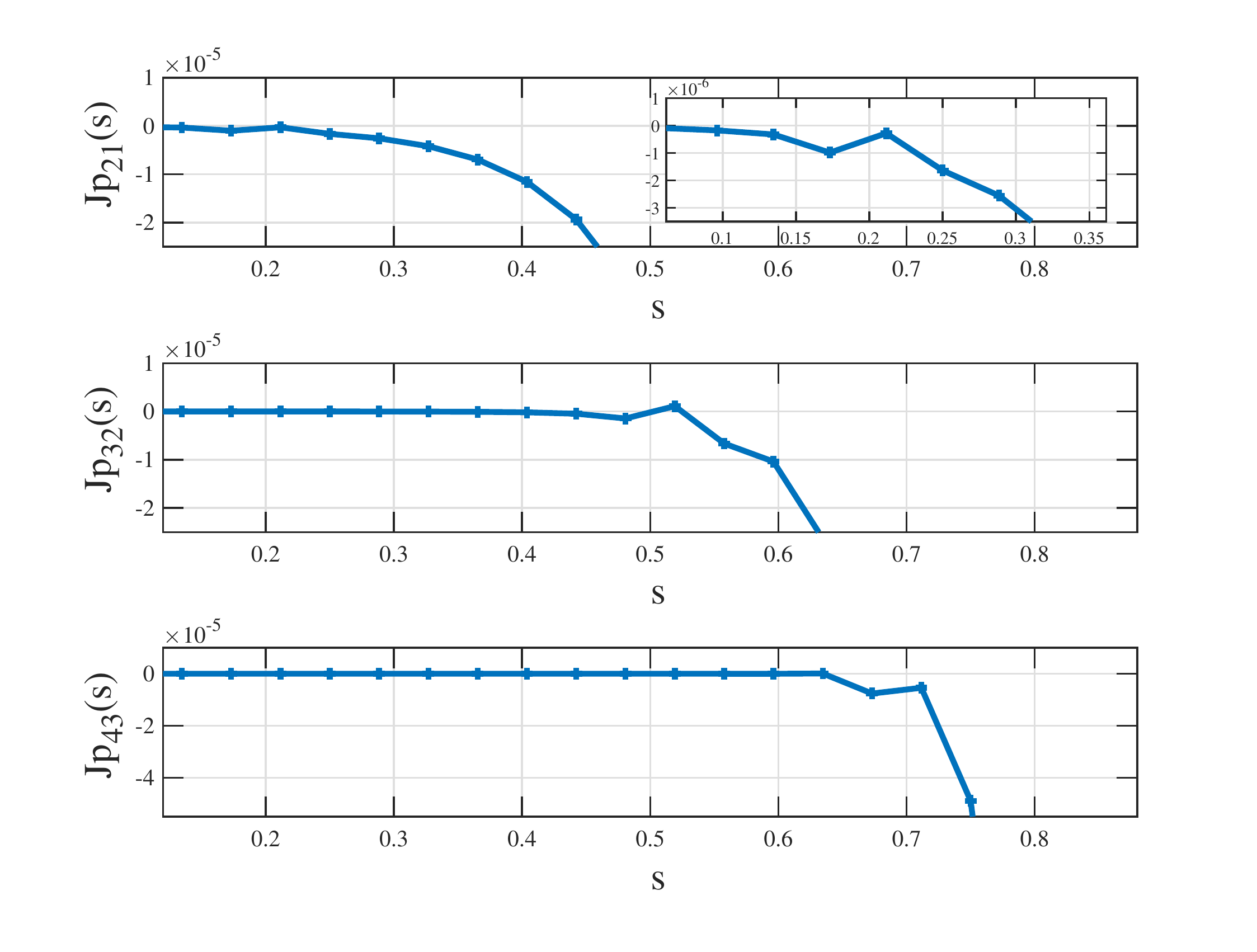}}
\parbox{6.0in}{
\caption{The (2,1) Fourier component $Jp_{21}(s)$ (top), the (3,2) Fourier component $Jp_{32}(s)$ (middle) and the (4,3) Fourier component $Jp_{43}(s)$ (bottom) of $Jp(s,u,v)$ as a function of the flux coordinate $s$ for a force-free slab equilibrium with three resonant surfaces. The rotational transform profile is $\iota(s) = 0.4+0.5s$, and the wall perturbation amplitudes are $\Delta_{20} = \Delta_{30} = \Delta_{40} = \Delta_{01} = \Delta_{02} = \Delta_{03} = 0.01$ and $\Delta_{21} = \Delta_{32} = \Delta_{43} = 0$.
\label{fig:threeJ1}} }
\end{center}
\end{figure}

\begin{figure}[ht]
\begin{center}
\resizebox{4.0in}{!}{\includegraphics{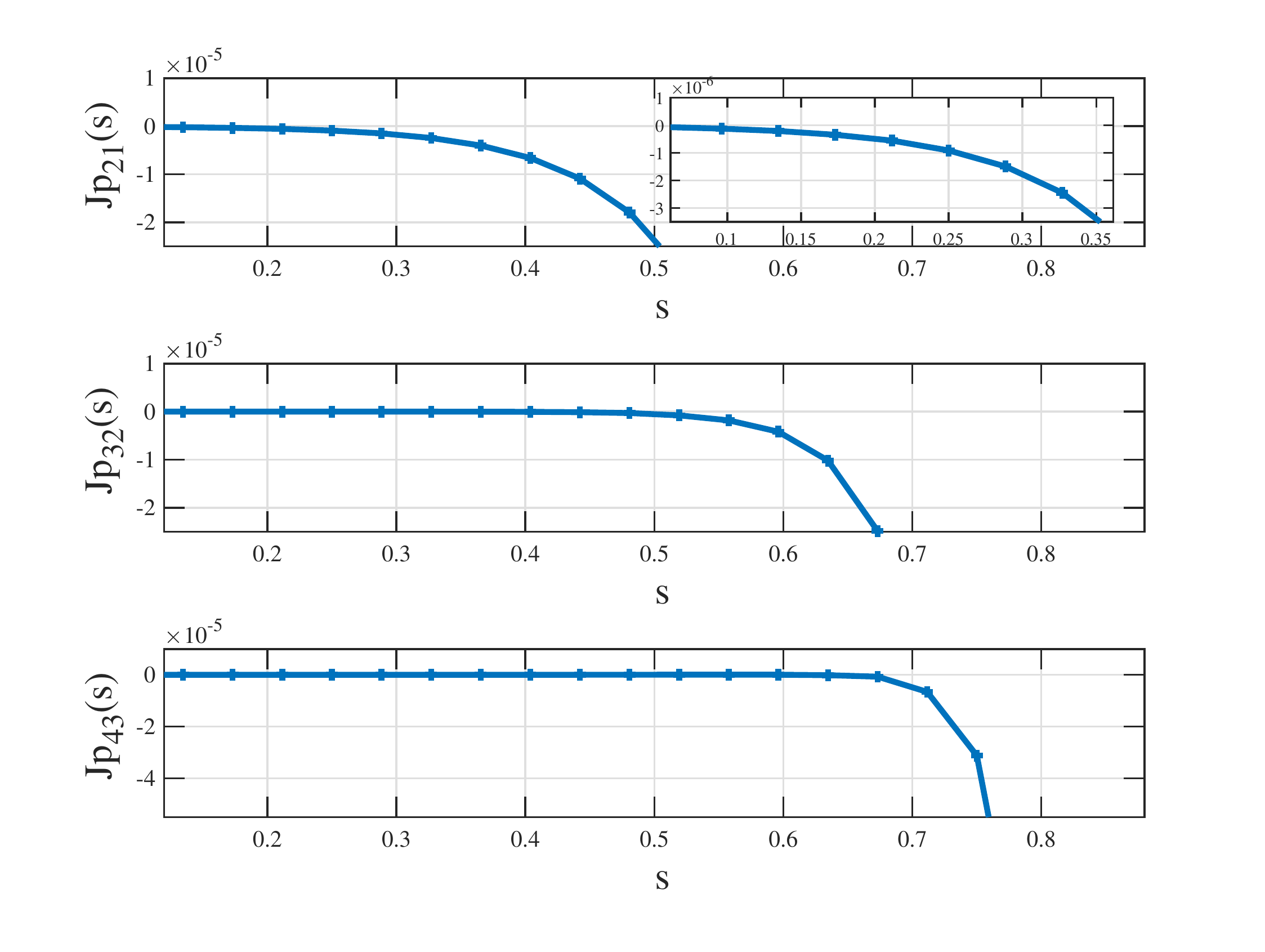}}
\parbox{6.0in}{
\caption{The (2,1) Fourier component $Jp_{21}(s)$ (top), the (3,2) Fourier component $Jp_{32}(s)$ (middle) and the (4,3) Fourier component $Jp_{43}(s)$ (bottom) of $Jp(s,u,v)$ as a function of the flux coordinate $s$ for a force-free slab equilibrium with three resonant surfaces. The rotational transform profile is $\iota(s) = 0.4+0.5s$, and the wall perturbation amplitudes are  $\Delta_{20} = \Delta_{30} = \Delta_{40} = \Delta_{01} = \Delta_{02} = \Delta_{03} = 0.01$, $\Delta_{21}=0.00083$, $\Delta_{32}=0.00222$, and $\Delta_{43}=0.00192$. Comparing these profiles to the parallel current profiles in \ref{fig:threeJ1}, we note that this choice of amplitudes for $\Delta_{21}$, $\Delta_{32}$, and $\Delta_{43}$ led to the elimination of the current singularities at the resonant surfaces.
\label{fig:threeJ2}} }
\end{center}
\end{figure}

Our final case is to consider three resonant surfaces, with a rotational transform $\iota(s)=0.4+0.5s$ admitting
the low-order rationals $1/2$, $2/3$, and $3/4$.
For the force-free $p'(s) = 0$ case shown in \Fig{fig:three1}, we set the boundary coefficients $\Delta_{20}=\Delta_{30}=\Delta_{40}=\Delta_{01}=\Delta_{02}=\Delta_{03}=0.01$.
There are singularities of $\psi_{21}$, $\psi_{32}$  and $\psi_{43}$ around the flux surfaces $s_{21} = 0.21154$, $s_{32} = 0.51923$ and 
$s_{43}= 0.71154$, respectively. In this case, the generalization of \Eqn{33} to the computation with three resonant surfaces is very slow to converge, and
we have employed an alternate strategy. We observe that a change of one wall harmonic, say $\Delta_{21}$, can have a significant effect on all three
singularities $D^2 \psi_{21}$, $D^2 \psi_{32}$, and $D^2 \psi_{43}$, so that the straightforward procedure that drives each wall harmonic
by its corresponding singular mode
in \Eqn{31} or \Eqn{33} can become ineffective. We therefore iterate on the coefficients by coupling their influence though a simple
version of a quasi-Newton procedure, setting
\be  \label{eq34}  
\left( \begin{array}{c}
       \Delta_{21}^{(n+1)} \\  \Delta_{32}^{(n+1)}  \\ \Delta_{43}^{(n+1)} 
    \end{array}  \right) = \left( \begin{array}{c}
       \Delta_{21}^{(n)} \\  \Delta_{32}^{(n)}  \\ \Delta_{43}^{(n)} 
    \end{array}  \right) - F^{-1} \left( \begin{array}{c}
       D^2 \psi_{21}^{(n)} \\  D^2 \psi_{32}^{(n)}  \\ D^2 \psi_{43}^{(n)}  \end{array} \right),
\ee
where $F$ is the $3 \times 3$ Jacobian 
$\partial(D^2 \psi_{21},D^2 \psi_{32},D^2 \psi_{43})/\partial (\Delta_{21},\Delta_{32},\Delta_{43})$
computed approximately via finite differences from separate NSLAB runs with varying wall perturbations. \Eqn{34} can also be regarded as the discretized form of a first
order ordinary differential equation in time that couples the dependence on the three wall
harmonics.
This procedure produces good values for the critical wall harmonics with only a few NSLAB runs, and we find the critical values
$\Delta_{21}=0.00083$, $\Delta_{32}=0.00222$, and $\Delta_{43}=0.00192$ as shown in \Fig{fig:three2}. 
In \Fig{fig:threeJ1} and \Fig{fig:threeJ2}, we show the profiles of the parallel currents corresponding to \Fig{fig:three1} and \Fig{fig:three2}, respectively. We again observe that the behaviors of the parallel current $Jp_{21}$ and $\psi_{21}$,  $Jp_{32}$ and $\psi_{32}$, and $Jp_{43}$ and $\psi_{43}$ are qualitatively similar. 

\section{Discussion}

We have developed a modified version of the MHD equilibrium and stability code NSTAB \cite{Tayl94} in a slab geometry that avoids complications
arising from the magnetic axis in toroidal geometries. We have used this code to study the possibility of using suitable wall modifications
to avoid the occurrence of singular current sheets that tend to arise at resonant flux surfaces where the rotational transform
assumes low-order rational values \cite{Gara99,BeMc88, Huds00,Nire19}. We find that a simple iterative procedure can be used to eliminate
one or two current sheets, while a more complicated procedure that takes additional mode coupling into effect 
suffices to remove three sheets. We have restricted our attention to relatively weak
current sheets that are generated by nonlinear interactions between ``sideband'' wall harmonics that can resonate with the fundamental harmonics
associated with the resonant flux surfaces. We have considered both force-free examples and examples with finite pressure gradients. Remarkably, finite pressure gradients at the resonant surfaces do not prevent us from removing the singularities at these surfaces, and do not affect the behavior of our solver. This could be an artifact of our focus on slab equilibria, although preliminary results in a toroidal geometry with the code NSTAB suggest otherwise, as we have found that we are also able to remove current singularities with finite pressure gradients in NSTAB stellarator equilibria. This is the subject of ongoing research with results to be reported in the near future.

\section*{Acknowledgements}

The authors thank H. Weitzner for suggesting this problem to us, and are grateful for many insightful discussions with D. Josell,  W. Sengupta, and H. Weitzner. The research of EK and AJC was partially supported by the U.S. Department of Energy, Office  of Science, Fusion Energy Sciences under Awards No. DE-FG02-86ER53223 and DE-SC0012398, and the Simons Foundation/SFARI (560651, AB).

\clearpage
\bibliographystyle{abbrv}
\bibliography{refs}

\begin{thebibliography}{10}

\bibitem{Bate80}
G.~Bateman.
\newblock {\em M{H}{D} Instabilities}.
\newblock MIT Press, Cambridge, MA, 1978.

\bibitem{BaBG78}
F.~Bauer, O.~Betancourt, and P.~Garabedian.
\newblock {\em A Computational Method in Plasma Physics}.
\newblock Springer-Verlag, New York, 1978.

\bibitem{BaBG84}
F.~Bauer, O.~Betancourt, and P.~Garabedian.
\newblock {\em Magnetohydrodynamic Equilibrium and Stability of Stellarators}.
\newblock Springer-Verlag, New York, 1984.

\bibitem{BBGW87}
F.~Bauer, O.~Betancourt, P.~Garabedian, and M.~Wakatani.
\newblock {\em The Beta Equilibrium, Stability and Transport Codes}.
\newblock Academic, Boston, 1987.

\bibitem{Beta88}
O.~Betancourt.
\newblock {BETAS}, a spectral code for three-dimensional magnetohydrodynamic
  equilibrium and nonlinear stability calculations.
\newblock {\em Communications on Pure and Applied Mathematics}, 41(5):551--568,
  1988.

\bibitem{Gara99}
O.~Betancourt and P.~Garabedian.
\newblock Numerical analysis of equilibria with islands in
  magnetohydrodynamics.
\newblock {\em Communications on Pure and Applied Mathematics}, 35, 1982.

\bibitem{BeMc88}
O.~Betancourt and G.~McFadden.
\newblock Nonparametric solutions to the variational principle of ideal
  magnetohydrodynamics.
\newblock {\em Lecture Notes in Pure and Applied Mathematics.}, 96:159--171,
  1985.

\bibitem{KAM}
N.~N. Bogoljubov, J.~A. Mitropoliskij, and A.~M. Samoilenko.
\newblock {\em Methods of Accelerated Convergence in Nonlinear Mechanics}.
\newblock Springer, Berlin, 2011.

\bibitem{Grad66}
H.~Grad.
\newblock Some new variational properties of hydromagnetic equilibria.
\newblock {\em The Physics of Fluids}, 7(8):1283--1292, 1964.

\bibitem{HiWh83}
P.~Hirshman and J.~Whitson.
\newblock Steepest-descent moment method for three-dimensional
  magnetohydrodynamic equilibria.
\newblock {\em The Physics of Fluids}, 26:3553S, 1983.

\bibitem{KrKu58}
M.~Kruskal and R.~Kulsrud.
\newblock Equilibrium of a magnetically confined plasma in a toroid.
\newblock {\em The Physics of Fluids}, 1:265, 1958.

\bibitem{Huds00}
J.~Loizu, S.~Hudson, A.~Bhattacharjee, and P.~Helander.
\newblock Magnetic islands and singular currents at rational surfaces in
  three-dimensional magnetohydrodynamic equilibria.
\newblock {\em Physics of Plasmas}, 22(2):022501, 2015.

\bibitem{Lortz1970}
D.~Lortz.
\newblock {\"U}ber die existenz toroidaler magnetohydrostatischer
  gleichgewichte ohne rotationstransformation.
\newblock {\em Zeitschrift f{\"u}r angewandte Mathematik und Physik ZAMP},
  21(2):196--211, Mar 1970.

\bibitem{Nire19}
M.~Mikhailov, J.~N{\"u}hrenberg, and R.~Zille.
\newblock Elimination of current sheets at resonances in three-dimensional
  toroidal ideal-magnetohydrodynamic equilibria.
\newblock {\em Nuclear Fusion}, 59(6):066002, 2019.

\bibitem{Tayl94}
M.~Taylor.
\newblock A high performance spectral code for nonlinear {MHD} stability.
\newblock {\em Journal of Computational Physics}, 110(2):407--418, 1994.

\bibitem{Weitzner14}
H.~Weitzner.
\newblock Ideal magnetohydrodynamic equilibrium in a non-symmetric topological
  torus.
\newblock {\em Physics of Plasmas}, 21(2):022515, 2014.

\end{thebibliography}

\end{document}